\documentclass[aps,prc,twocolumn,superscriptaddress,floatfix,nofootinbib]{revtex4-1}

\usepackage{threeparttable}
\usepackage{graphicx}
\usepackage{adjustbox}
\usepackage{dcolumn}
\usepackage{tensor}
\usepackage{bm}
\usepackage{float}
\usepackage{color}
\usepackage[usenames,dvipsnames]{xcolor}
\usepackage{listings}
\usepackage{fancyvrb}
\usepackage{slashed}
\usepackage{amsmath}
\usepackage{amssymb}
\usepackage{amsfonts}
\usepackage{enumitem}
\usepackage{listings}
\usepackage{mathtools}
\usepackage{slashed}
\usepackage{verbatim}
\usepackage{hyperref}
\usepackage[margin=1.0in]{geometry}
\usepackage{slashed}
\usepackage{braket}
\usepackage{dcolumn}
\usepackage[normalem]{ulem}
\usepackage{cancel}
\usepackage{makecell}
\newcommand{\cev}[1]{\reflectbox{\ensuremath{\vec{\reflectbox{\ensuremath{#1}}}}}}

\usepackage{accents}
\usepackage[normalem]{ulem} 
\makeatletter
\DeclareRobustCommand{\cev}[1]{%
  \mathpalette\do@cev{#1}%
}
\newcommand{\do@cev}[2]{%
  \fix@cev{#1}{+}%
  \reflectbox{$\m@th#1\vec{\reflectbox{$\fix@cev{#1}{-}\m@th#1#2\fix@cev{#1}{+}$}}$}%
  \fix@cev{#1}{-}%
}
\newcommand{\fix@cev}[2]{%
  \ifx#1\displaystyle
    \mkern#23mu
  \else
    \ifx#1\textstyle
      \mkern#23mu
    \else
      \ifx#1\scriptstyle
        \mkern#22mu
      \else
        \mkern#22mu
      \fi
    \fi
  \fi
}

\makeatother

\def\ee{\end{eqnarray}}

\newcommand{\be}{\begin{eqnarray}}
\newcommand{\en}{\end{eqnarray}}
\newcommand{\bea}[1]{\left(\begin{array}{#1}}
\newcommand{\ena}{\end{array}\right)}
\newcommand{\ba}{\begin{eqnarray}}
\newcommand{\ea}{\end{eqnarray}}

\newcommand\lrnab{\raise .8ex\hbox{$^\leftrightarrow$} \hspace{-8.8pt}
\nabla}
\newcommand\lnab{\raise .8ex\hbox{$^\leftarrow$} \hspace{-9.8pt}
\nabla}
\newcommand\rnab{\raise .8ex\hbox{$^\rightarrow$} \hspace{-9.8pt}
\nabla}

\begin{document}


\title{The Gallium Neutrino Absorption Cross Section and its Uncertainty}

\author{S. R. Elliott}
\affiliation{Los Alamos National Laboratory, Los Alamos, NM 87545, USA}
\author{V. N. Gavrin}
\affiliation{Institute for Nuclear Research of the Russian Academy of Sciences, Moscow 117312, Russia} 
\author{W. C. Haxton}
\affiliation{Department of Physics, University of California, Berkeley, CA 94720, USA}
\affiliation{Lawrence Berkeley National Laboratory, Berkeley, CA 94720, USA}
\author{T. V. Ibragimova}
\affiliation{Institute for Nuclear Research of the Russian Academy of Sciences, Moscow 117312, Russia} 
\author{E. J. Rule}
\affiliation{Department of Physics, University of California, Berkeley, CA 94720, USA}

\date{\today}

\begin{abstract}
In the recent Baksan Experiment on Sterile Transitions (BEST), a suppressed rate of neutrino absorption on a gallium target was observed,
consistent with earlier results from neutrino source calibrations of the SAGE and GALLEX/GNO solar neutrino experiments.  The BEST collaboration,
utilizing a 3.4 MCi $^{51}$Cr neutrino source, found
observed-to-expected counting rates at two very short baselines of $R=0.791 \pm 0.05$ and $0.766 \pm 0.05$, respectively.
Among recent neutrino experiments, BEST is notable for the simplicity of both its neutrino spectrum, 
line neutrinos from an electron-capture source whose intensity can be measured to a estimated precision of 0.23\%, and its absorption cross section,
where the precisely known rate of electron capture to the gallium ground state, $^{71}$Ge$(e^-,\nu_e)^{71}$Ga(g.s.), establishes a minimum value. 
However, the absorption cross section uncertainty is a common systematic in the BEST, SAGE, and GALLEX/GNO
neutrino source experiments.  Here we update that cross section, considering a variety of electroweak corrections and the role of transitions to
excited states, to establish both a central value and reasonable uncertainty,  thereby enabling a more accurate assessment of
the statistical significance of the gallium anomalies.  Results are given for $^{51}$Cr and  $^{37}$Ar sources. The revised neutrino capture rates are used in a re-evaluation of the
BEST and gallium anomalies.
\end{abstract}

\pacs{}

\maketitle

\section{Introduction: The Ga neutrino anomaly}
\begin{figure*}[ht]   
\centering
\includegraphics[scale=0.4]{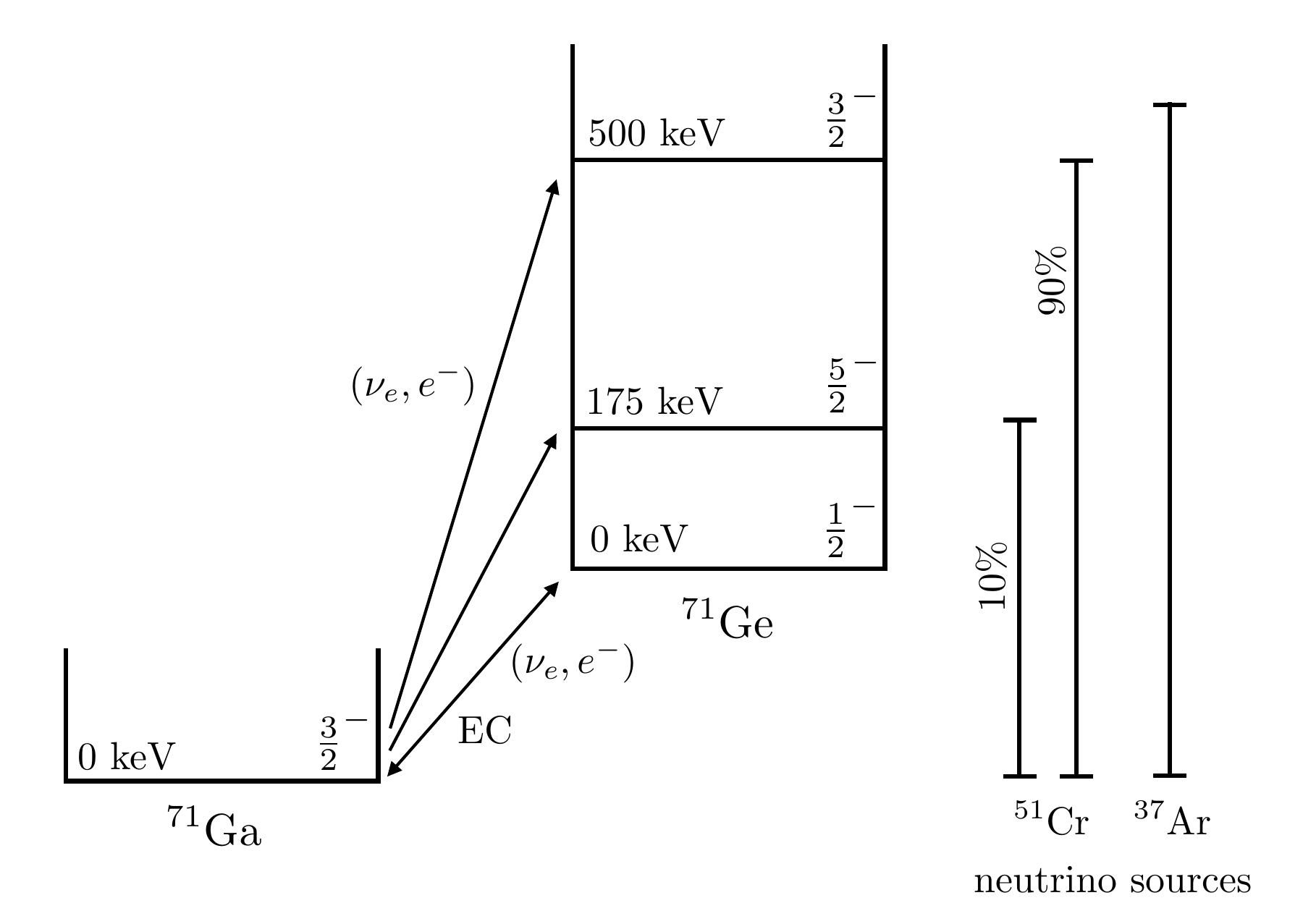}
\caption{Level diagram for $^{71}$Ga$(\nu_e,e^-)^{71}$Ga showing the states that contribute
to the absorption of $^{51}$Cr and $^{37}$Ar EC neutrinos.}
\label{fig:excited}
\end{figure*}

\noindent
The possibility of additional, very weakly interacting ``sterile" neutrinos, beyond the three light neutrinos of the standard model, has
been raised frequently in the literature \cite{Acero2022,Abazajian2012,Gariazzo2015,Giunti2019,Boser2020,Diaz2020,Seo2020,Dasgupta2021,Acero2022}.
They arise naturally in extensions of the standard model that account for
nonzero neutrino masses.   Sterile neutrinos have been discussed in connection with the LSND experiment,
the reactor neutrino anomaly, the SAGE and GALLEX/GNO neutrino calibration experiments,
and with efforts to reconcile oscillation parameters derived from T2K and NOvA  \cite{Acero2022,Seo2020,deGouvea2022}.   \\
~\\
\noindent
In the radiochemical SAGE and GALLEX/GNO solar neutrino experiments, a large mass of Ga (30-50 tons) was exposed to the solar neutrino flux for 
a period of about a month, during which neutrino capture occurs via $^{71}$Ga$(\nu_e,e^-)^{71}$Ge.  The produced atoms of radioactive $^{71}$Ge,
$\tau_{1/2} =11.43~ \pm$ 0.03 d, 
were then chemically extracted and counted as they decay back to  $^{71}$Ga via electron capture.  These experiments established
capture rates that, in combination with those from the chlorine and Kamioka experiments, indicated a pattern of solar neutrino fluxes
that could not be easily reconciled with solar models, helping to motivate a new generation of solar neutrino detectors: Super-Kamiokande \cite{SK}, the Sudbury Neutrino
Observatory \cite{SNO}, and Borexino \cite{Borexino1,Borexino2}.  This led to the discovery of neutrino mass and oscillations and the detection of an energy-dependent distortion of the solar neutrino flux,
reflecting the interplay between vacuum and matter-enhanced oscillations \cite{review1,review2}.\\
~\\
While both gallium experiments utilized tracers to demonstrate
the reliability of the chemical extraction, direct cross checks on their overall efficiencies for neutrino detection were also performed.  Intense 
$^{51}$Cr and $^{37}$Ar electron-capture (EC) neutrino sources of known strength
were placed at the center of the Ga targets, and the additional production of $^{71}$Ge measured.   Four such calibrations \cite{Ansel1995,Ab1999,Hampel1998,Ab2006} were performed, which when combined
yield a ratio of the observed to expected counting rates of $R =0.866 \pm 0.054$.  The discrepancy between this result and $R=1$ is known as the
gallium anomaly.\\
~\\
The gallium anomaly and other short-baseline neutrino discrepancies motivated the recent 
Baksan Experiment on Sterile Transitions (BEST) \cite{BEST1,BEST2}.  BEST, employing an exceptionally intense 3.4MCi 
$^{51}$Cr neutrino source, measured the rate of neutrino reactions at two distances by dividing the Ga target reactor into inner and outer volumes.
This opened up the possibility of detecting an oscillation signal.  While no distance dependence was seen,
the counting rates were again well below expectations, with $R=0.791 \pm 0.05$ and $0.766 \pm 0.05$ for the inner (shorter baseline)
and outer volumes, respectively. \\
~\\
A critical issue in the analysis of BEST and earlier Ga neutrino source experiments is the cross section for $^{71}$Ga$(\nu_e,e^-)^{71}$Ge, as this is a common systematic
in these measurements.  For $^{51}$Cr and $^{37}$Ar neutrino sources, the contributing transitions from the $^{71}$Ga ground state are to
the ground state and first two excited states of $^{71}$Ge, as shown in Fig. \ref{fig:excited}.  
The Ga anomaly cannot be attributed entirely to uncertainties in the neutrino cross section, due to the dominance of the 
strong ${3 \over 2}^- \rightarrow {1 \over 2}^-$  transition
to the $^{71}$Ge ground state, as this transition strength is precisely determined by the known EC rate of $^{71}$Ge.   Even if only this
contribution is included, a $\sim 2 \sigma$ discrepancy remains.   In addition, two
allowed (GT) transitions to $^{71}$Ge excited states, the ${5 \over 2}^-$ and ${3 \over 2}^-$ levels at 175 and 500 keV,
respectively, also contribute to the total $^{51}$Cr neutrino absorption cross section.  The contributions of these transitions have generally been deduced 
from surrogate probes of Gamow-Teller (GT) strength, forward-angle (p,n) or ($^3$He,t) scattering --- despite long-established concerns about the 
reliability of these probes when applied to specific weak transitions \cite{HH1,HH2}. \\
~\\
While one cannot attribute the Ga anomaly entirely to nuclear physics, the central value and uncertainty of the cross section can
influence the statistical significance of the BEST result, its possible interpretation in terms of new physics, and its consistency with other tests of
neutrino properties.   The purpose of this letter is to 1) re-examine the relationship between the $^{71}$Ge(g.s.)($e^-,\nu_e)^{71}$Ga(g.s.) electron capture rate and 
the $^{71}$Ga(g.s.)$(\nu_e,e^-)^{71}$Ge(g.s.) 
cross section, in order to deduce the best value and uncertainty of the latter;  and 2) reconsider the excited-state contributions in light of new data testing the
proportionality between (p,n) or ($^3$He,t) cross sections and experimentally known weak rates.   In 1), we examine (or re-examine) 
several $\approx 1\%$ corrections that can impact the proportionality between the $\mathrm{g.s.}\leftrightarrow \mathrm{g.s.}$ inverse reactions.  In 2), our focus is on
defining a reasonable uncertainty for the excited state contribution, based on a critical examination of the reliability of
such surrogate interactions as probes of specific weak GT transitions.\\

\section{The $^{71}$Ge electron capture rate}
\noindent
One would like to derive from the known electron capture rate for $^{71}$Ge the strength of the ground-state GT 
transition of the inverse neutrino reaction cross section.   In addition to the half life \cite{Hampel1985,DataSheet},
\begin{equation}
 \tau_{1 \over 2}[^{71}\mathrm{Ge}] = 11.43 \pm 0.03~ \mathrm{d} ,
 \end{equation}
relevant experimental 
information includes the $Q_\mathrm{EC}$ value for the decay \cite{Frekers2016}, the difference in the atomic masses
\begin{eqnarray}
 Q_\mathrm{EC} &=& M[^{71}\mathrm{Ge}]-M[^{71}\mathrm{Ga}] \nonumber \\
 &=&232.443 \pm 0.093 \mathrm{~keV} ,
 \label{eq:Q}
 \end{eqnarray}
and the $P_K$, $P_L$, and $P_M$ electron-capture probabilities and associated atomic binding energies \cite{Ab1999b},
\begin{eqnarray}
P_K &=& 0.88,~~~~~~ E_\mathrm{bind}=10.37 ~\mathrm{keV}\nonumber \\
P_L &=& 0.103,~~~~~E_\mathrm{bind}=1.2 ~\mathrm{keV}\nonumber \\
P_M &=& 0.017,~~~~~E_\mathrm{bind}=0.12~\mathrm{keV}.
\end{eqnarray}
The $^{71}$Ge $\rightarrow$ $^{71}$Ga electron capture rate can then be written 
\begin{widetext}
\begin{equation}
 \omega = {\mathrm{ln}[2] \over \tau_{1 \over 2} } = {G_F^2 \cos^2{\theta_C} \over 2 \pi} ~|\phi_{1s}|_\mathrm{avg}^2 ~ E_{\nu,1s}^2~ { \textstyle \left[2(1+\epsilon_o^{1s}) (1+ {P_L+P_M \over P_K} )\right]}~ ~ g_A^2~ [2 ~\mathrm{B^{(\nu,e)}_{GT}}(\mathrm{gs})]~[1+g_{v,b}]_\mathrm{EC} ~[ 1 + \epsilon_q]  
 \label{eq:sigmao}
 \end{equation}
 \end{widetext}
 
\noindent
The various terms appearing above are as follows:
\begin{enumerate}[wide, labelwidth=!, labelindent=0pt]
\item {\it Q value:} The neutrino energy $E_{\nu,1s}$ for electron K capture.  Neglecting a very small nuclear recoil correction, it is given by the energy constraint
\begin{eqnarray}
 Q_\mathrm{EC}&=& E_{\nu,1s} + 10.37 \mathrm{~keV} \nonumber \\
 \Rightarrow E_{\nu,1s}  &=& 222.1 \pm 0.1 \mathrm{~keV} \nonumber
\end{eqnarray}
\item {\it Branchings:} The factor $\left[2(1+\epsilon_o^{1s}) (1+ {P_L +P_M \over P_K}) \right]$ relates the total capture rate to the rate for capture of a single $1s$ electron, with the 
contributions of $L$ and $M$ capture included through use of the experimentally known branching ratios.  This procedure requires the introduction of a
rearrangement (or overlap-exchange) correction $\epsilon_0^{1s}$
to account for the imperfect overlap of the state created by annihilating a $1s$ electron in the $^{71}$Ge atomic ground state, with states appropriate 
for the Coulomb field of $^{71}$Ga.   That is, while the instantaneous annihilation of the $1s$ electron in $^{71}$Ge will lead dominantly to a virtual state that
decays by emitting $K$-capture Auger electrons and X-rays, atomic rearrangement generates small contributions from $L$ and $M$ capture.
Similar corrections would be needed for other channels.  That is, the total rate would be proportional to
\begin{widetext}
\begin{eqnarray}
 \sum_i |\phi_i |_\mathrm{avg}^2 E_{\nu,i}^2 (1+\epsilon_o^i ) &=&  |\phi_{1s}|_\mathrm{avg}^2 ~ E_{\nu,1s}^2~ (1+\epsilon_o^{1s}) \left[ 1 + { \sum_{i\ne 1s} |\phi_i |_\mathrm{avg}^2 E_{\nu,i}^2 (1+\epsilon_o^i ) \over |\phi_{1s} |_\mathrm{avg}^2 E_{\nu,1s}^2 (1+\epsilon_o^{1s} )}  \right]  \nonumber \\
&=& |\phi_{1s}|_\mathrm{avg}^2 ~ E_{\nu,1s}^2~ (1+\epsilon_o^{1s}) {\textstyle \left[ 1 + {P_L+P_M \over P_K}  \right] }
\label{eq:ps}
\end{eqnarray}
\end{widetext}
where $|\phi_{i}|_\mathrm{avg}^2$ is the $K$, $L$, or $M$ atomic density at the nucleus, $E_{\nu,i}$ is the associated energy of the emitted neutrino, and
$\epsilon_o^i$ is the overlap and exchange correction needed in the $i$th channel \cite{Benois1950,Odiot1956,Bahcall62,Vatai70}.   Bahcall \cite{Bahcall62} noted that such corrections to theory were needed
to reproduce precise experimental $L/K$ capture ratios and estimated their sizes.  As he has emphasized, if the theoretical expression on the left-hand side of Eq. (\ref{eq:ps}) is used, the
inclusion of the $\epsilon_i$ would have little net impact, as these factors diminish $P_K$ but enhance $P_L$ and $P_M$.  Here, however, we make use of the experimentally
measured probabilities $P_K$, $P_L$ and $P_M$ to write the total rate in terms of the $1s$-capture rate, so inclusion of
overlap/exchange correction for the $1s$ channel is needed.\\
~\\
The values for $\epsilon_o^{1s}$ given by Bahcall \cite{Bahcall62,Bahcall63a,Bahcall63b,Bahcall65} and by Vatai \cite{Vatai70} are -0.018 and 
-0.008, respectively.  Though the correction is small, there is a relatively large fractional difference between the results.  (In contrast, their corrections for  $L$ (0.083 and 0.088, respectively) and $M$ 
(0.247 and 0.188, respectively) capture are in better agreement,
with fractional differences of 6\% and 31\%, respectively).  We adopt as a nominal value the average and take twice the standard deviation as the 95\% confidence level (C.L.), yielding
$\epsilon_o^{1s} = -0.013 \pm 0.014$ and 
\[(1+ \epsilon_o^{1s}) (1+ {P_L+P_M \over P_K})= 1.122 \pm 0.016.\]
\item {\it Weak couplings:} We adopt Particle Data Group (PDG) values for the Fermi constant, $G_F/(\hbar c)^3 = 1.1664 \times 10^{-5}/$GeV$^2$, and Cabibbo angle,
$\cos{\theta_C}  =0.9743$, and the PERKEO III value \cite{Perkeo2019} for the axial vector coupling $g_A=1.2764$. (The PERKEO III experiment employed a novel pulsed cold neutron
source to greatly reduce systematic uncertainties, yielding a result that is both exceptionally precise and statistics dominated.  The PDG value for $g_A$ employs an
error-bar inflation of 2.7 to account for the scatter among past experiments, thereby eroding the impact of the new technique.)

Note, however, that the choice of weak couplings and their uncertainties do not influence our results.  As all transition rates are taken from experiment, any change in the weak couplings 
would be absorbed into the fitted BGT value.   Weak coupling uncertainties --- whether taken from the PDG or elsewhere -- are too small to influence the overall error budget
of our cross section calculations.
\item {\it Electron density at the nucleus:}  $|\phi_{1s}|^2_\mathrm{avg}$ is the $^{71}$Ge $1s$ atomic density at the nucleus.  The nuclear amplitudes for the EC transitions of interest involve convolutions of the GT operator ---
the space-like component of the nuclear axial current --- with leptonic wave functions, 
\[ \langle j_f  | \int  d {\boldsymbol r}~ \phi_{\nu_e}^*({\boldsymbol r})  \phi_{1s}({\boldsymbol r}) \sum_{i=1}^A {\boldsymbol{\sigma}}(i) \tau_-(i) ~ \delta({\boldsymbol r} - {\boldsymbol r}_i)~ | j_i  \rangle \]
where we abbreviate the nuclear ground states of $^{71}$Ge and $^{71}$Ga as $|j_i \rangle$ and $|j_f \rangle$, respectively.
As $q_\nu R_N \ll 1$, where $R_N$ is the nuclear radius and $q_\nu$ is the magnitude of the neutrino's three-momentum, one can approximate the neutrino plane wave
within the nucleus by $\phi_{\nu_e}^*({\boldsymbol r}) \sim 1$.  (The leading correction to this approximation will be evaluated below.)  Similarly, given that the atomic wave
function varies slowly over the nuclear scale, $ \phi_{1s}({\boldsymbol r})$ can be removed from the integral and replaced by an average value.
Most commonly  $|\phi_{1s}|_\mathrm{avg}$ is computed by folding the electron probability density
with the normalized $^{71}$Ge proton charge distribution, then integrating over the nuclear volume.\\
~\\
In his 1997 work \cite{Bahcall97}, Bahcall used three relativistic, self-consistent Hartree-Fock calculations that took into
account the finite extent of the nucleus, the Breit interaction, vacuum polarization, and self-energy corrections, averaging the resulting wave functions over
the nuclear volume to obtain $|\phi_{i} |_\mathrm{avg} ^2$ for $K$, $L$, and $M$ capture.    The calculations, performed by three independent groups,
agreed at the $\pm 0.2\%$ level.   We are not aware of any subsequent calculations that
are as complete.  While \cite{Bahcall97} includes references to the atomic methods employed by the three groups, the $^{71}$Ge results
were provided as private communications and are not described in separate publications.   As the relationship between the dimensionless numerical quantity given
in \cite{Bahcall97} and the density $|\phi_{1s}|_\mathrm{avg}$ may not be obvious to readers, we provide some of the needed definitions here.\\
~\\
The dimensionless quantity evaluated in \cite{Bahcall97}, given by the quantity in square brackets below, is related to the dimensionful quantities
we define on the left below by
\begin{equation}
 \sum_i |\phi_i |_\mathrm{avg}^2 E_{\nu,i}^2 = {(m_ec^2)^5 \over (\hbar c)^3} {1 \over 4 \pi}  \left[ \sum_i  g_i^2 \tilde{E}_{\nu,i}^2  \right]
 \label{eq:three}
 \end{equation}
 where $\tilde{E}_{\nu,i} \equiv {E}_{\nu,i}/m_ec^2$.  The factor of $1/4 \pi$ appears because Bahcall evaluated the $s$-wave radial density, not the full density.
 As the square-bracketed quantity depends on $Q_\mathrm{EC}$,  a small correction is needed because $Q_\mathrm{EC} =232.69$ keV was used in \cite{Bahcall97}, while the current value is given by Eq. (\ref{eq:Q}).  Plugging
 in the numerical values from \cite{Bahcall97} one finds
 \begin{eqnarray}
 & & \left[ \sum_i  g_i^2 \tilde{E}_{\nu,i}^2  \right]^{Q_\mathrm{EC}=232.443~\mathrm{keV}} \nonumber \\ &\approx & 0.9978   \left[ \sum_i  g_i^2 \tilde{E}_{\nu,i}^2  \right]^{Q_\mathrm{EC} = 232.69~\mathrm{keV}} \nonumber \\
  &=& 0.01440
  \label{eq:num}
  \end{eqnarray}
 We again group terms so that we can use experimental EC ratios, finding from Eqs. (\ref{eq:three}) and (\ref{eq:num})
 \begin{eqnarray}
&& {(m_ec^2)^5 \over (\hbar c)^3} ~{0.01440 \over 4 \pi} \nonumber \\
&=&  |\phi_{1s} |_\mathrm{avg}^2 E_{\nu,1s}^2 \left[1 + {1 + \epsilon_o^{1s}  \over 1 + \epsilon_o^L} {P_L \over P_K} + {1 + \epsilon_o^{1s}  \over 1 + \epsilon_o^M} {P_M \over P_K}  \right] \nonumber \\
&=& |\phi_{1s} |_\mathrm{avg}^2 E_{\nu,1s}^2 \left\{  \begin{array}{cc}  1.121, & \mathrm{Bahcall}  \\ 1.123, & \mathrm{Vatai} \end{array} \right.~~~
 \label{eq:dens}
 \end{eqnarray}
 where the overlap and exchange factors arise because, following Eq. (\ref{eq:ps}), canceling terms are implicitly included in
 the probabilities $P_K$, $P_L$, and $P_M$.   We find the the result depends only weakly on whether we take these corrections from  \cite{Bahcall62} or from \cite{Vatai70}.
 Evaluating this expression yields
 \begin{equation}
 ~~~~(\hbar c)^3 |\phi_{1s} |_\mathrm{avg} ^2 =  (7.21 \pm 0.03) \times 10^{-4} \mathrm{~MeV}^3 
 \label{eq:dens2}
 \end{equation}
The uncertainty is determined from the standard deviations of the three
atomic calculations reported in \cite{Bahcall97} and of the overlap and the exchange corrections of 
Bahcall and Vatai.  These are combined in quadrature, then doubled to give the 95\% C.L. range given in Eq. (\ref{eq:dens2}).  This 
procedure thus takes into account differences apparent from the spread among competing calculations,
but not those that could arise if the calculations being compared employed common but flawed
assumptions.\\
~\\
One can recast this numerical result in terms of a more familiar density, 
the Schr\"{o}dinger density for an electron bound to a point charge $Z$, evaluated at the origin.   One finds
\begin{equation}
(\hbar c)^3 |\phi_{1s} |_\mathrm{avg} ^2 = R {(Z \alpha m_e c^2)^3 \over \pi}\Big|_{Z=32}
\end{equation}
where Eq. (\ref{eq:dens2}) determines the numerical proportionality factor, $R= 1.333$. 

\item B$_\mathrm{GT}$ {\it convention:} In this paper, all B$_\mathrm{GT}$ values are given for the neutrino reaction direction, $^{71}$Ga$(\nu_e,e^-)^{71}$Ge.
For the $\mathrm{gs} ~\rightarrow~ \mathrm{gs}$ EC direction,
\begin{eqnarray}
&&\mathrm{B}_\mathrm{GT}^\mathrm{EC}(\mathrm{gs}) \nonumber \\
&=& {1 \over 2j_i +1} \left| \langle j_f={\textstyle {3 \over 2}} || \sum_{1=1}^A \boldsymbol\sigma(i) \tau_-(i) ||j_ i={\textstyle {1 \over 2}} \rangle \right|^2 \nonumber  \\
&=&  {2 \over 2j_i +1} \left| \langle j_f={\textstyle {1 \over 2}} || \sum_{1=1}^A \boldsymbol\sigma(i) \tau_+(i) ||j_ i={\textstyle {3 \over 2}} \rangle \right|^2 \nonumber \\
& \equiv & 2 ~\mathrm{B^{(\nu,e)}_{GT}}(\mathrm{gs})  
\end{eqnarray}
Thus, $\mathrm{B^{EC}_{GT}}(\mathrm{gs})$ is given as  2$\mathrm{B^{(\nu,e)}_{GT}}(\mathrm{gs})$ in Eq. (\ref{eq:sigmao}).

\item {\it Weak magnetism correction:}  $[1 + \epsilon_q]$ is the correction to $B_\mathrm{GT}$ arising from contributions beyond the allowed approximation.   
Because of the very low momentum transfer, these corrections
are expected to be small and dominated by the interference term between the GT
amplitude and weak magnetism.  This interference generates a term linear in the three-momentum transfer. 
We find a correction to the GT transition probability of
\begin{equation}
\epsilon_q = {2 E_{\nu,1s} \over 3 m_N g_A} \left[ \mu_{T=1} + {\langle {3 \over 2}^- || \sum_{i=1}^A  \boldsymbol\ell(i) \tau_-(i) || {1 \over 2}^- \rangle \over \langle {3 \over 2}^- || \sum_{i=1}^A  \boldsymbol\sigma(i) \tau_-(i) || {1 \over 2}^- \rangle } \right] 
 \label{eq:forb}
 \end{equation}
where $m_N$ is the nucleon mass and $\mu_{T=1} \, \approx \, 4.706$ is the isovector magnetic moment.   As the spin
contribution to weak magnetism is effectively determined by the EC capture rate, only the orbital contribution must be
taken from theory.  \\
~\\
While the large isovector magnetic moment makes the weak magnetism correction relatively insensitive to nuclear structure
uncertainties, one still must estimate the orbital contribution.  We do this using the shell model (SM), retaining all Slater determinants 
within the $2p_{3/2}1f_{5/2}2p_{1/2}1g_{9/2}$ model space, and 
employing three effective interactions designed for this space,
GCN2850 \cite{GCN2850}, jj44b \cite{jj44b}, and JUN45 \cite{JUN45}.   \\
~\\
We selected these interactions because of the extensive
literature comparing their predictions to experiment, specifically, how well they reproduce measured moments,
transitions, and low-lying nuclear spectra.   For example, in \cite{JUN45} comparisons are made to experiment for
binding energies, magnetic and quadrupole moments, B(E2) values, and nuclear spectra of a large set of $2p_{3/2}1f_{5/2}2p_{1/2}1g_{9/2}$ nuclei, 
including both $^{71}$Ga and $^{71}$Ge.   In the paper presenting the jj44b interaction \cite{jj44b},
the properties and spectroscopy of odd isotopes of Ga (including $^{71}$Ga) were used as test of its quality.
Side-by-side comparisons of JUN45 and jj44b predictions for
spectra, quadrupole moments, and B(E2) values for the even isotopes of Ge are made in \cite{JGH2012,Mane}, and for the odd-isotopes of Ga (including $^{71}$Ga) in
\cite{Sri}.   The literature on GCN2850 predictions is somewhat more limited: the interaction has been employed in studies of weak
process like $\beta \beta$ decay ($^{76}$Ge) and WIMP scattering ($^{73}$Ge).  Representative work includes \cite{Klos,Caurier,Menendez}.\\
~\\
The dimension of the SM space for $^{71}$Ge is about $1.5 \times 10^8$. The diagonalizations  were performed
with the Lanczos-algorithm code BIGSTICK \cite{Johnson:2013bna,Calvin}.  We found
\[{\textstyle \langle {3 \over 2}^- }|| \sum_{i=1}^A  \boldsymbol\ell(i) \tau_-(i) ||{\textstyle {1 \over 2}^- \rangle }= \left\{ \begin{array}{ll} 0.48 & \mathrm{GCN2850} \\ 0.69 & \mathrm{jj44b} \\ 0.005 & \mathrm{JUN45} \end{array} \right.\]
so that the ratio that enters in Eq. (\ref{eq:forb}) is
\[{{\textstyle \langle {3 \over 2}^- }|| \sum_{i=1}^A  \boldsymbol\ell(i) \tau_-(i) ||{\textstyle {1 \over 2}^- \rangle } \over {\textstyle \langle {3 \over 2}^- }|| \sum_{i=1}^A  \boldsymbol\sigma(i) \tau_-(i) ||{\textstyle {1 \over 2}^- \rangle }}_\mathrm{exp}= \left\{ \begin{array}{ll} -0.81 & \mathrm{GCN2850} \\ -1.18 & \mathrm{jj44b} \\ -0.01 & \mathrm{JUN45} \end{array}\right.\]
where the magnitude of the GT matrix element is taken from experiment, while the relative sign is the SM prediction.  From the average and standard deviation of these theory results, we find
\[ {{\textstyle \langle {3 \over 2}^- }|| \sum_{i=1}^A  \boldsymbol\ell(i) \tau_-(i) ||{\textstyle {1 \over 2}^- \rangle } \over {\textstyle \langle {3 \over 2}^- }|| \sum_{i=1}^A  \boldsymbol\sigma(i) \tau_-(i) ||{\textstyle {1 \over 2}^- \rangle }}_\mathrm{exp}= -0.7 \pm 1.2~~(95\% ~\mathrm{C.L.})  \]
where the assigned uncertainty is again twice the standard deviation.
Because the large isovector magnetic moment dominates Eq. (\ref{eq:forb}), the estimate of the forbidden corrections is relatively stable, despite substantial
differences in the SM estimates of the orbital angular momentum matrix element.  The end result
\[  \epsilon_q =(4.9 \pm 1.5)   \times 10^{-4} ~~~~ (95\%~\mathrm{C.L.})\]
shows that the weak magnetism correction is negligible.
We have also evaluated this correction using the full momentum dependence of the weak transition amplitude,
doing a standard multipole expansion, obtaining a result consistent with the above to the precision shown.\\

\item {\it Radiative corrections:} The factor $[1 + g_{v,b}]_\mathrm{EC}$ is the EC radiative correction.  Past work has either explicitly \cite{Bahcall97} or implicitly assumed that radiative corrections
would affect the electron capture rate and the inverse neutrino capture cross section similarly, and thus would be
effectively included in calculations that extract an effective GT matrix element from electron capture, then use that amplitude
in computing the inverse $(\nu_e,e^-)$ reaction.  Sirlin \cite{Sirlin} has pointed out that certain single-nucleon short-range contributions
to radiative corrections are universal.  But other contributions, notably bremsstrahlung, affect electron capture and neutrino
reactions unequally, with the differences dependent on the $Q$ value of the reaction \cite{Kurylov}.   When we 
evaluate the corresponding corrections for neutrino capture $[1 + g_{v,b}]_{(\nu,e)}$, we will obtain a ratio
of radiative corrections that isolates the non-universal contribution, which we will then evaluate.
\end{enumerate} 
~\\
Collecting all of the results from this section and utilizing the $11.43 \pm 0.03$ d (1$\sigma$) half life of $^{71}$Ge we find
\begin{eqnarray}
 \omega &=& (7.019 \pm 0.037 ) \times 10^{-7}/s ~~~~~(95\%~\mathrm{C.L.})\nonumber \\ 
 &=& (8.122 \pm 0.122)  \times 10^{-6}  ~\mathrm{B^{(\nu,e)}_{GT}}(\mathrm{gs}) ~ [1+ g_{v,b}]_\mathrm{EC} \nonumber
 \end{eqnarray}
 and therefore
 \begin{eqnarray}
&\mathrm{B}^{(\nu,e)}_\mathrm{GT}(\mathrm{gs})[1+ g_{v,b}]_\mathrm{EC}  \equiv \tilde{\mathrm{B}}^{(\nu,e)}_\mathrm{GT}(\mathrm{gs})\nonumber\\
&= 0.0864 \pm 0.0013 ~~~(95\%~\mathrm{\,C.L.})
 \label{eq:result}
 \end{eqnarray}
where the various uncertainties noted above have been combined in quadrature.   The ground-state transition probability
is conservatively known to a precision of about 1.5\%:  the primary motivation for the detailed discussion above was to establish
this uncertainty.  Our recommended best value of 0.0864 is consistent with most past estimates of this quantity, e.g., 
0.087 \cite{HH1}, 0.0863 \cite{Bahcall97}, and 0.0864 \cite{Semenov2020} --- though the agreement is a bit fortuitous,
arising because differences in rate components conspire to cancel.
~\\

\section{The $^{71}$Ga$(\nu_e,e^-)^{71}$Ge ground state cross section}
\noindent
The $^{71}$Ga$(\mathrm{gs})(\nu_e,e^-)^{71}$Ge$(\mathrm{gs})$ neutrino capture cross section can be written in terms of $ \tilde{\mathrm{B}}^{(\nu,e)}_\mathrm{GT}(\mathrm{gs})$,
\begin{widetext}
\be
 \sigma_{\mathrm{gs}}  = {G_F^2 \cos^2{\theta_C} \over  \pi}  ~ p_e E_e~{\cal F}(Z_f,E_e)  ~ g_A^2~ \tilde{\mathrm{B}}^{(\nu,e)}_\mathrm{GT}(\mathrm{gs}) ~{[1+g_{v,b}]_{(\nu,e)} \over [1+g_{v,b}]_\mathrm{EC} }~[ 1 + \epsilon_q] 
 \label{eq:sigmaor}
 \ee
 \end{widetext}
Nuclear recoil has been neglected as the target mass $M_T \gg E_e$.
We evaluate the cross section for electron-capture neutrinos produced by $^{51}$Cr and $^{37}$Ar, for which the contributing lines are listed in Table \ref{tab:neutrino_capture}.
 \begin{table*}
 \caption{Neutrino source parameters and various correction factors that enter into the calculation of the cross section $\sigma_\mathrm{gs}(\nu_e+^{71}$Ga$\rightarrow e^-+^{71}$Ge$)$ for $^{51}$Cr and $^{37}$Ar neutrino sources. We report the energy of the incoming neutrino $E_\nu$, the corresponding neutrino branching ratio, and the energy of the final-state electron $E_e$. $\mathcal{F}(Z_f,E_e)$ is the Coulomb factor of the electron, obtained by combining the various correction factors $F_0$, $L_0$, $U$, and $S$ (see text). Finally, $\epsilon_q$ governs the strength of the forbidden corrections to the GT amplitude, and $[1+g_{v,b}]_{(\nu,e)}/[1+g_{v,b}]_\mathrm{EC}$ measures the difference in radiative corrections that enter into the calculation of the respective cross sections for neutrino capture and the inverse process of electron capture.}
 \label{tab:neutrino_capture}
 \centering
 {\renewcommand{\arraystretch}{1.6}
 \begin{ruledtabular}
 \begin{tabular}{ccccccccccc}
 Source & $E_{\nu}$ (MeV) & Branching & $E_e$ (MeV) & $F_0$ & $L_0$ & $U$ & S & $\mathcal{F}(Z_f,E_e)$ & $1+\epsilon_q$ (95\% C.L.) & $\frac{[1+g_{v,b}]_{(\nu,e)}}{[1+g_{v,b}]_\mathrm{EC}}$\\
\hline
 $^{51}$Cr & 0.7524 & 0.0140 & 1.031 & 2.791 & 1.0034 & 0.9986 & 0.9920 & 2.774 & $1.0034\pm 0.0010$ & 0.995\\
           & 0.7518 & 0.0842 & 1.030 & 2.791 & 1.0034 & 0.9986 & 0.9920 & 2.775 & $1.0034\pm 0.0010$ & 0.995\\
           & 0.7465 & 0.8025 & 1.025 & 2.795 & 1.0035 & 0.9986 & 0.9920 & 2.779 & $1.0034\pm 0.0010$ & 0.995\\
           & 0.4323 & 0.0015 & 0.711 & 3.335 & 1.0053 & 0.9985 & 0.9876 & 3.306 & $1.0017\pm 0.0005$ & 0.997\\
           & 0.4317 & 0.0092 & 0.710 & 3.338 & 1.0053 & 0.9985 & 0.9876 & 3.309 & $1.0017\pm 0.0005$ & 0.997\\
           & 0.4264 & 0.0886 & 0.705 & 3.360 & 1.0053 & 0.9985 & 0.9874 & 3.330 & $1.0017\pm 0.0005$ & 0.997\\
 $^{37}$Ar & 0.8138 & 0.0111 & 1.092 & 2.750 & 1.0031 & 0.9986 & 0.9925 & 2.734 & $1.0037\pm 0.0011$ & 0.995\\
           & 0.8135 & 0.0866 & 1.092 & 2.750 & 1.0031 & 0.9986 & 0.9925 & 2.734 & $1.0037\pm 0.0011$ & 0.995\\
           & 0.8107 & 0.9023 & 1.089 & 2.752 & 1.0031 & 0.9986 & 0.9925 & 2.736 & $1.0037\pm 0.0011$ & 0.995\\
 \end{tabular}
 \end{ruledtabular}
 }
 \end{table*}
The neutrino energies $E_\nu$ and the corresponding branching ratios are computed from the respective $Q$ values in $^{51}$Cr and $^{37}$Ar, 752.4 and 813.9 keV, the $K$-shell binding energies, 5.99 and 3.21 keV, the $L$-shell binding energies, 0.70 and 0.33 keV, the $M$-shell binding energies, 0.074 and 0.029 keV,
the $K$/$L$/$M$ branching ratios of 0.891/0.094/0.016 and 0.902/0.0866/0.011, and the 9.93\% branching
ratio for $^{51}$Cr to decay to the first excited $\textstyle{5 \over 2}^-$ state in $^{51}$V at 321.1 keV.\\
~\\
The various terms in Eq. (\ref{eq:sigmaor}) are:
\begin{enumerate}[wide, labelwidth=!, labelindent=0pt]
\item {\it Kinematics:} The energy and three-momentum magnitude of the outgoing electron are denoted $E_e$ and $p_e$, respectively.  For the neutrino reactions
of interest off $^{71}$Ga 
\[ E_e = E_\nu - Q_\mathrm{EC} +m_e -0.09 \mathrm{~keV}  \]
where $Q_\mathrm{EC}$ in given in Eq. (\ref{eq:Q}).  We follow Bahcall \cite{Bahcall97} in including a very small 0.09 keV correction
for the energy lost to electronic re-arrangement, as the electron cloud adjusts to the nuclear charge change.   
For transitions to the ${\textstyle {5 \over 2}^-}$(175 keV) and ${\textstyle {3 \over 2}^-}$(500 keV) excited states in $^{71}$Ge, the 
nuclear excitation energies would be added to $Q_\mathrm{EC}$.
\item {\it Coulomb corrections:} ${\cal F}(Z_f,E_e)$ corrects the phase space for the Coulomb distortion of the outgoing electron plane wave.  Following \cite{Wilkinson}, \cite{Hayen}, and \cite{Behrens} this
correction is decomposed as follows
\begin{eqnarray}
&&{\cal F}(Z_f,E_e)\nonumber \\
&=& F_0(Z_f,E_e)~ L_0(Z_f,E_e)~U(Z_f,E_e)~S(Z_f,E_e) \nonumber
\end{eqnarray} 
with
\begin{eqnarray}
F_0(Z_f,E_e) = 4 (2 p_e R_N)^{2(\gamma-1)} e^{\pi y}  {|\Gamma(\gamma+i y) |^2 \over (\Gamma(1+2 \gamma))^2}~~~ \nonumber\\
\gamma \equiv \sqrt{1-(\alpha Z_f)^2} ~~~~~~~~~~
y \equiv \alpha Z {E_e \over p_e} ~~~~~~~~~~~~~\nonumber
\end{eqnarray}
$F_0$ is taken from the solution of the Dirac equation for an electron of momentum $p_e$ in a point Coulomb potential generated
by a charge $Z_f$, with $Z_f=32$ here.  This correction is kept finite by its evaluation at the nuclear surface, often taken to be 
$R_N \, \approx \,1.2 A^{1/3}$ fm and interpreted as the edge of a nucleus of uniform density. 
We fix the $^{71}$Ge r.m.s. charge radius to 4.05 fm,  the average of the charge radii for $^{70}$Ge and $^{72}$Ge, as measured in electron scattering \cite{Devries}, then use the relationship
for a nucleus of uniform density to determine
\[    R_N =\sqrt{5 \over 3} \sqrt{ \langle r^2 \rangle } =5.23 \mathrm{~fm}  \, \approx \, 1.263 A^{1/3}~\mathrm{fm}\big|_{A=71} \]
which we use in the evaluation.  This initial estimate then must be corrected:
\begin{enumerate}
\item $L_0$ accounts for most effects of the finite charge distribution.  For a nucleus with a uniform density and thus a sharp surface at $R_N$,
the Dirac solution can be continued to the origin by numerically integrating.  We take $L_0$ from the tables of Behrens and Janecke \cite{Behrens}, who performed the
integration for $R_N=1.2 A^{1/3}$ fm.  We adjusted that result to account for the difference between this estimate of the r.m.s. charge radius and the experimental value used here, using Eq. (16) of \cite{Hayen} (or Eq. (2)
of \cite{Wilkinson90}). 
\item The factor $U(Z_f,E_e)$ represents the difference between the Coulomb distortion computed for a uniform charge distribution
and that resulting from the use of a more realistic Fermi distribution with an equivalent r.m.s. radius.  We use the parameterization of Wilkinson \cite{Wilkinson}, 
also recently discussed in \cite{Hayen} (see Eqs. (29) and (30)).  
\item $S(Z_f,E_e)$ is a correction for atomic screening within the nuclear volume, which we take from Rose \cite{Rose}.  A
comparison of various prescriptions for the atomic screening correction is presented in \cite {Hayen} (see Fig. 5 in this reference), showing generally good agreement, except
for very low $E_e \lesssim 1.1 m_e$.
\end{enumerate}
 ~~\\
Table \ref{tab:neutrino_capture} gives the Coulomb factors $\mathcal{F}(Z_f,E_e)$ and the constituent corrections $F_0$, $L_0$, $U$, and $S$ for $^{71}$Ga$(\nu_e,e^-)^{71}$Ge for $^{51}$Cr and $^{37}$Ar neutrino sources.
 
\item {\it Weak magnetism:} $[1+\epsilon_q]$ corrects for the omission of forbidden contributions, dominated
in this case by the interference between the GT amplitude and weak magnetism.  
 After integrating over electron angles, the correction linear in weak magnetism takes on
a  form identical to Eq. (\ref{eq:forb})
\begin{eqnarray}
 &\epsilon_q& = {2  \over 3 m_N g_A} \left(E_\nu + E_e -{m_e^2 \over E_e} \right)\nonumber\\
 &\times& \left[ \mu_{T=1} + {\langle {3 \over 2}^- || \sum_{i=1}^A  \boldsymbol\ell(i) \tau_-(i) || {1 \over 2}^- \rangle \over \langle {3 \over 2}^- || \sum_{i=1}^A  \boldsymbol\sigma(i) \tau_-(i) || {1 \over 2}^- \rangle}_\mathrm{exp}  \right] 
 \label{eq:forbr}
 \end{eqnarray}
 apart from the kinematic factor.  The resulting forbidden correction $\epsilon_q$ is shown in Table \ref{tab:neutrino_capture}.
The $2\sigma$ uncertainty reflects the differences among the three SM estimates of the orbital
matrix element, as discussed previously.

\item {\it Non-universal radiative correction:} The ratio
\[ { [1+g_{v,b}]_{(\nu,e)} \over [1+g_{v,b}]_\mathrm{EC} }\]
accounts for the difference between the radiative corrections \cite{Sirlin}  for 
neutrino absorption and those for electron capture (contained in $\mathrm{\tilde{B}_{GT}}(\mathrm{gs})$).
While in a given low-energy weak nuclear process the radiative correction can be significant
(few \% \cite{Kurylov}), it is frequently assumed \cite{Bahcall97} that these corrections
 affect inverse reactions $(e^-,\nu_e)$ and $(\nu_e,e^-)$ similarly, and thus are implicitly included 
 when the $(\nu_e,e^-)$ nuclear transition matrix elements are determined from
 known electron-capture rates.   Were this the case, the ratio above would be 1.
 However, while this universality assumption holds for charge-current reactions
 producing electrons/positrons only in the final or only in the initial state, Kurylov et al. \cite{Kurylov} have shown that it is not preserved
 in the comparison between electron capture and $(\nu_e,e^-)$.\\
 ~\\
 \noindent
 Kurylov et al. \cite{Kurylov} evaluated the one-nucleon $W\gamma$-loop and bremsstrahlung
 contributions to the radiative corrections (Figs. 1a and 3 of \cite{Kurylov}), finding that the bremsstrahlung
 contribution breaks the universality due to its dependence on the $Q$ value.  While the calculation treats the electron in $(e^-,\nu_e)$ as a free state, the results evaluated for $E_e \rightarrow m_e$ 
 should approximate those needed for the weakly bound electrons of interest here.  (The radiative corrections
 describe short-range loops and radiation associated with the strong Coulomb field near the nucleus.  
 The bound electron wave function varies over atomic scales, not nuclear ones,  providing justification for this assumption.  
 In \cite{Czarnecki}, similar issues are discussed in comparing muonium decay with free muon decay.)  \\
   ~\\
 The results shown in the last column of Table \ref{tab:neutrino_capture} were derived using Eqs. (4), (5), (51), and (52) of \cite{Kurylov}.
  The difference in the one-nucleon/bremsstrahlung contributions to electron capture (implicitly 
  absorbed into $\tilde{\mathrm{B}}_{GT}(\mathrm{gs})$) and neutrino reactions yields a
  correction to the neutrino absorption cross section of $\approx$ 0.5\%.\\
  ~ \\
  In addition to the effects discussed above, there are nucleus-dependent radiative corrections --- contributions involving 
  more than one nucleon (Fig. 1b of \cite{Kurylov}) as well as the nuclear Green's function corrections to terms treated in leading order as one-nucleon
  contributions.  Such corrections for the axial current have not yet been estimated and thus are not included here.
\end{enumerate}
 ~\\
  {\it Cross sections:} Combining all of the results above yields
  \begin{equation}
   \sigma_{\mathrm{gs}} = \left\{ \begin{array}{cc} (5.39 \pm 0.08) \times 10^{-45} \mathrm{~cm}^2  &~~~ {}^{51}\mathrm{Cr} \\ (6.45 \pm 0.10) \times 10^{-45} \mathrm{~cm}^2 & ~~~   {}^{37}\mathrm{Ar} \end{array} \right.
   \label{eq:gs}
   \end{equation}
at $95\%$ C.L.

\section{Excited-state contributions}
\noindent
The excited-state contributions, which we will find increase the total cross section by about 6\%, 
can also be extracted from experiment, specifically from forward-angle (p,n) scattering.   However, past work
has either failed to employ an appropriate effective interaction, or failed to propagate associated experimental and theoretical
uncertainties -- raising questions about the reliability of the extracted GT strengths.  In this section we 
describe an improved extraction that yields both the needed transition strengths and reasonable estimates of their uncertainties.\\
~\\
The potential importance of excited-state contributions was noted by
Kuzmin \cite{Kuzmin1966}, when he proposed $^{71}$Ga as a solar neutrino detector in 1966.
In fact, one of the motivations for the $^{51}$Cr and $^{37}$Ar source experiments is that they populate the same excited states -- the 5/2$^-$ and 3/2$^-$ states at 175 and 500 keV --
that contribute to $^7$Be solar neutrino capture (see Fig \ref{fig:excited}).\\
~\\
In his 1978 Ga cross section study, Bahcall \cite{Bahcall1978} used systematics to constrain the excited-state contributions,
identifying transitions in neighboring nuclei that might be similar; that is, na\"ively of a $2p_{3/2} \leftrightarrow 1f_{5/2}$ character.  Bahcall identified nine $3/2^- \rightarrow 5/2^-$ transitions
of known strength with log(ft) values ranging from 5.9 to 7.5, and consequently assigned log(ft)$\gtrsim 6$ to the transition to the 175 keV state in $^{71}$Ge.  Similarly,
he found eight $3/2^- \rightarrow 3/2^-$ transitions with log(ft) values ranging from 5.0 to 5.8, assigning log(ft)$\gtrsim 5$ to the transitions to the 500 and 710 keV states.
Using these bounds, Bahcall argued that the excited-state contribution to the $^{51}$Cr absorption cross section
would be $\lesssim$ 14.6\%.   But the potential fallibility of such arguments was 
pointed out in \cite{HH1}, as there are exceptions to these patterns in neighboring nuclei.\\
~\\
Alternatively, one might attempt a microscopic calculation of the strengths of the excited-state transitions.  Indeed, SM
calculations of the B$_\mathrm{GT}$ values for exciting the 175 and 500 keV states were performed early on by Baltz et al. \cite{Baltz1984} and by Mathews et al. \cite{Mathews1985}.
But even today --- as we will describe later --- this is a dubious undertaking due to the weakness of these transitions.  In the allowed approximation, the transition probabilities are proportional
to the B$_\mathrm{GT}$ value
\begin{eqnarray}
\mathrm{B_{GT}}({j_i\alpha_i \rightarrow j_f \alpha_f}) &=& {1 \over 2j_i+1} | M_\mathrm{GT}|^2 \nonumber \\
&=& {1 \over 2 j_i +1} |\langle j_f \alpha_f || \hat{O}_\mathrm{GT}^{J=1} || j_i \alpha_i \rangle |^2  \nonumber\\
\hat{O}_\mathrm{GT}^{J=1} &\equiv& \sum_{j=1}^A {\boldsymbol \sigma}(j) \tau_+(j)
\end{eqnarray}
where ${\boldsymbol \sigma}$ is the Pauli spin matrix, $\tau_+$ is the isospin raising operator, $||$ denotes a matrix element reduced in angular momentum,
and $j_i \alpha_i$ and $j_f \alpha_f$ denote the quantum numbers of the initial and final states, respectively, with the angular momentum $j$ made explicit.    
From the known EC rate for $^{71}$Ge and from the lower bounds Bahcall used for the transitions to the 175 and
500 keV levels, one finds
\begin{eqnarray}
\label{eq:BGTa}
\mathrm{B_{GT}} \left(^{71}\mathrm{Ga} ~\mathrm{gs} \rightarrow  {}^{71}\mathrm{Ge} ~ 175 \mathrm{~keV} \right) &\lesssim& 0.004\nonumber\\
\mathrm{B_{GT}} \left(^{71}\mathrm{Ga} ~\mathrm{gs} \rightarrow  {}^{71}\mathrm{Ge} ~ 500 \mathrm{~keV}  \right) &\lesssim& 0.04
\label{BahcallGT}
\end{eqnarray}
values smaller than the ground-state B$_\mathrm{GT}$ value of Eq. (\ref{eq:result}).
As the total $\mathrm{B}_\mathrm{GT}$ strength, summed over all final states, is given approximately by the Ikeda sum rule $3(N-Z) = 27$ \cite{Ikeda1964}, we see that the transitions to the 5/2$^-$ and 3/2$^-$ states
exhaust less than $\approx$ 0.01\% and $0.1\%$ of the sum-rule strength, respectively.  Consequently, one expects calculations to be sensitive to wave-function details,
including the interactions used, the adopted SM spaces, etc.  A weak transition typically indicates
substantial interferences among the individual amplitudes in the transition density matrix. Indeed, early attempts to estimate the needed excited-state contributions to the cross section, using the SM \cite{Baltz1984,Mathews1985},
schematic effective interactions, and truncated model spaces, yielded results that varied by orders of magnitude, depending on the specific simplifications adopted.\\
~\\
Here we make use of the full power of the modern SM -- carefully tuned interactions like those discussed in the previous section, and the ability to treat all Slater determinants in the
$2p_{3/2}1f_{5/2}2p_{1/2}1g_{9/2}$ shell -- but only to estimate corrections that typically alter results at the level of $\lesssim$ 10\%.  Apart from these corrections,
the needed weak GT strengths are extracted from experiment.\\
~\\
Specifically, the possibility that excited-state GT strength could be measured through surrogate reactions, (p,n) or ($^3$He,t), 
generated significant interest in the solar neutrino community.  The approximate proportionality between medium-energy forward-angle (p,n) cross sections
and nuclear B$_\mathrm{GT}$ profiles is well established \cite{Taddeucci}.  This method was applied to the $^{71}$Ga transitions of interest by Krofcheck et al. \cite{Krofcheck}.
From (p,n) measurements at 120 and 200 MeV they deduced
\begin{eqnarray}
\label{eq:BGT}
\mathrm{B_{GT}^{(p,n)}} \left(^{71}\mathrm{Ga}~\mathrm{gs} \rightarrow  {}^{71}\mathrm{Ge} ~ 175 \mathrm{~keV} \right) &\lesssim& 0.005 \nonumber \\
\mathrm{B_{GT}^{(p,n)}} \left(^{71}\mathrm{Ga}~\mathrm{gs} \rightarrow  {}^{71}\mathrm{Ge} ~ 500 \mathrm{~keV}  \right) &=& 0.011 \pm 0.002 \nonumber \\
& & 
\label{eq:Krofcheck}
\end{eqnarray}
results qualitatively consistent with Bahcall's expectations based on systematics.  However, the use of this method
in the case of weak transitions can be problematic, as described in \cite{HH1,HH2}.  Here we extend these previous analyses with the
goal of better quantifying the excited-state contributions to the neutrino absorption cross section.\\
~\\
The same transitions were studied using ($^3$He,t) at 420 MeV \cite{FrekersGa}.  This method can achieve higher resolution,
but has been applied less frequently to the light nuclei we will later use to test the reliability of charge-exchange mappings of Gamow-Teller strength.
The results are
\begin{eqnarray}
\label{eq:BGTb}
\mathrm{B_{GT}^{(p,n)}} \left(^{71}\mathrm{Ga}~\mathrm{gs} \rightarrow  {}^{71}\mathrm{Ge} ~ 175 \mathrm{~keV} \right) &=& 0.0034\pm0.0026 \nonumber \\
\mathrm{B_{GT}^{(p,n)}} \left(^{71}\mathrm{Ga}~\mathrm{gs} \rightarrow  {}^{71}\mathrm{Ge} ~ 500 \mathrm{~keV} \right) &=& 0.0176 \pm 0.0014 \nonumber \\
& &
\label{eq:Frekers}
\end{eqnarray}
As the tension between Eqs. \ref{eq:Krofcheck} and \ref{eq:Frekers} for the transition to the $\textstyle{3 \over 2}^-$ 
state exceeds $3\sigma$, we will treat the two data sets separately, rather than combining them.\\ 
~\\
The work in \cite{HH1,HH2} exploited the empirical observation \cite{Watson} that the effective operator for forward-angle (p,n) 
scattering includes a subdominant contribution from a tensor operator $\hat{O}^{J=1}_\mathrm{T}$,
\begin{eqnarray}
M^\mathrm{(p,n)} &\equiv& M_\mathrm{GT} + \delta M_\mathrm{T}, ~~~~ M_\mathrm{T} \equiv \langle j_f \alpha_f ||  \hat{O}^{J=1}_\mathrm{T}  || j_i \alpha_i \rangle\nonumber\\
\hat{O}^{J=1}_\mathrm{T}&=&\sqrt{8 \pi}  \sum_{j=1}^A  \left[Y_2(\Omega_j) \otimes \boldsymbol\sigma(j) \right] _{J=1} \tau_+(j)
\label{eq:MT}
\end{eqnarray}
where $\delta \, \approx \, 0.1$, so that
\be
\mathrm{B_{GT}^{(p,n)}} = {1 \over 2 j_i +1} | \langle j_f \alpha_f || M^\mathrm{(p,n)} || j_i \alpha_i \rangle|^2~ .
\ee
The need for the tensor correction in forward-angle scattering, where the momentum transfer is minimal and thus the interactions can
occur at long range, should not be a surprise:  the central part of the one-pion-exchange potential generates a target response proportional to $M_\mathrm{GT}$ while
the tensor part generates $M_\mathrm{T}$.   In cases where $M_\mathrm{GT}$ is weak but $M_\mathrm{T}$ is strong, $M^\mathrm{(p,n)}$ will be
an unreliable probe of B$_\mathrm{GT}$ strengths.  An example where this would be the case is an $\ell$-forbidden M1 transition, where the dominant amplitude links orbitals with the
quantum numbers $[n,\ell,j+\textstyle{1 \over 2}]$ and $[n-1,\ell+2,j=\ell+\textstyle{3 \over 2}]$.  Such transitions are often found at low energy in nuclear spectra,
as a consequence of an approximate pseudospin symmetry \cite{Ginocchio}.  A candidate $\ell$-forbidden transition \cite{HH1} is $^{71}$Ga(3/2$^-$) $\rightarrow^{71}$Ge(5/2$^-$),
which would be described in the na\"ive SM as $1f_{5/2}$ (n hole) $\rightarrow$ $2p_{3/2}$ (p particle).  State-of-the-art SM studies performed
here and in another recent study \cite{Kostensalo} show that the transition density matrix does have an important
$\ell$-forbidden component.   \\
~\\
The analysis in \cite{HH1} estimated $\delta$ by examining GT transitions in $1p$- and $2s1d$-shell nuclei, but did not evaluate the experimental and
theoretical errors in the determination, nor how they would propagate into an estimate of $\delta$ and consequently the $^{71}$Ga excited-state cross section.  Given
the BEST anomaly, it is now important to do so.  The data examined in \cite{HH1} were sensibly chosen, involving mirror transitions where
$\beta$ decay and (p,n) transition strengths were both available from experiment, including transitions near closed shells where levels are well
separated and thus their SM wave functions less sensitive to small changes in effective interactions.  However, we have made some changes in the data set, reflecting
new information that has become available.  We also assess theoretical uncertainties by employing several available effective interactions in computing $M_\mathrm{T}$:
to relate $M^\mathrm{(p,n)}$ and $M_\mathrm{GT}$, $|M_\mathrm{T}|$ and the sign of $M_\mathrm{T}/M_\mathrm{GT}$ must be computed. \\
~\\
The data we use in determining $\delta$ are given in Table \ref{tab:data} (compare to Table 1 of \cite{HH1}) and consist primarily of isospin mirror transitions
where both (p,n) and $\beta$-decay strengths are known.  Eight of these cases are taken from the compilation
of $\mathrm{B_{GT}^{(p,n)}}$ of \cite{Watson}.   To convert the proportionality between (p,n) scattering and  $\beta$ decay into an equivalence, a normalization must
be introduced.  Often this is done by using, for each target nucleus, a strong $\beta$ decay transition, which can still be problematic
if there are corrections due to $\hat{O}_\mathrm{T}^{J=1}$ that affect normalizing transitions in differential ways.  The study of \cite{Watson} 
instead computed normalizing cross sections in the distorted-wave impulse approximation, employing a phenomenological interaction fitted to a large body of data.  This
then avoids the issue of nucleus-by-nucleus normalization systematics.\\
~\\

\begin{table*}[ht]
\caption{ \label{tab:data} Weak transitions and their beta decay and (p,n) amplitudes deduced from experiment.  The sign of $M_\mathrm{GT}$ has been taken as positive.  $M_\mathrm{T}$ and its sign relative to $M_\mathrm{GT}$ have been taken from theory, except
 for the case of the sign of $^{32}$S (see text).}
  {\renewcommand{\arraystretch}{1.8}
 \begin{adjustbox}{width=1\textwidth}
\begin{tabular}{ldcccccc}
\hline
\hline
 Reaction\footnote{Transitions are between ground states unless otherwise specified.} & \multicolumn{1}{c}{~~~~~~~~~log(ft)\footnote{Taken from the ENSDF compilations.}}&~ ($2J_i+1$)B$_\mathrm{GT}$~& $\textstyle{\mathrm{B_{GT}^\mathrm{(p,n)}} \over \mathrm{B_{GT}}}$  & $M_\mathrm{GT}$  &  $M^\mathrm{(p,n)}$  &  $M_\mathrm{T}$\footnote{$2s1d$-shell uncertainties (excluding $^{39}$K$\rightarrow ^{39}$Ca) correspond to the $1\sigma$ spread of matrix elements computed from the USDA, USDB \cite{USD}, and Brown-Wildenthal \cite{Wildenthal} interactions. All other $M_T$ uncertainties are assumed to be 10\%.} & ${M_\mathrm{T} \over M^\mathrm{(p,n)}}$ \\
\hline
~${}^{13}$C$(\textstyle{1 \over 2}^-$) $\leftrightarrow \, {}^{13} $N$(\textstyle{1 \over 2}^-$) & 3.6648(5) & 0.404 $\pm$ 0.002& 1.85$\pm$ 0.22 & 0.636$\pm$0.002 & 0.87$\pm$0.05  & $2.8\pm 0.3$ & 3.2\\
$\begin{array}{l}  {}^{14}\mathrm{C}(\textstyle{0}^+) \\   {}^{14}\mathrm{O}(\textstyle{0}^+) \end{array} \Big\} \rightarrow  {}^{14}\mathrm{N}(1^+\,3.95\,\mathrm{MeV})$  & 3.131(17) & 2.79 $\pm$ 0.11 & 0.97$\pm$0.12 & 1.67$\pm$0.03 &  1.65$\pm$0.10  & $0.086\pm 0.009$ & 0.052\\
~${}^{15}$N$(\textstyle{1 \over 2}^-$) $\leftrightarrow \,^{15}$O$(\textstyle{1 \over 2}^-$) & 3.6377(8) & 0.509$\pm$0.003 & 2.04$\pm$0.24 & 0.713$\pm$0.002 &  1.02$\pm$0.06  &  $3.3\pm 0.3$ & 3.2\\
~${}^{17}$O$(\textstyle{5 \over 2}^+$) $\leftrightarrow \,^{17}$F$(\textstyle{5 \over 2}^+$)  & 3.3562(5) & 6.280$\pm$0.011 & 0.91$\pm$0.11  & 2.506$\pm$0.002 &  2.39$\pm$0.14 & $1.7\pm 0.2$ & 0.69\\
~${}^{18}$O$(\textstyle{0}^+$) $\leftrightarrow \,^{18}$F$(\textstyle{1}^+$)  & 3.5700(19) & 3.045$\pm$0.013 & 1.12$\pm$0.13   & 1.745$\pm$0.003 & 1.84$\pm$0.11 &  -0.04$\pm$0.03 & -0.02\\
$\begin{array}{l}  {}^{18}\mathrm{O}(\textstyle{0}^+) \\  {}^{18}\mathrm{Ne}(\textstyle{0}^+)  \end{array} \Big\} \rightarrow {}^{18}\mathrm{F} (1^+\,1.70\,\mathrm{MeV})$ & 4.470(15) & 0.128$\pm$0.004 & 1.33$\pm$0.17 & 0.358$\pm$0.006 &  0.41$\pm$0.025  & 0.8$\pm$0.3 & 2\\
~${}^{19}$F$(\textstyle{1 \over 2}^+$) $\leftrightarrow \,^{19}$Ne$(\textstyle{1 \over 2}^+$)  & 3.2329(24) & 3.184$\pm$0.024 & 1.29$\pm0.15$  & 1.784$\pm$0.007 &  2.02$\pm$0.12 & 0.08$\pm$0.03 & 0.04 \\
~${}^{19}$F$(\textstyle{1 \over 2}^+$) $\leftrightarrow \,^{19}$Ne$(\textstyle{3 \over 2}^+$~1.55\,MeV)  & 5.71(5) & 0.0294$\pm$0.0034 & 2.65$\pm0.41$  & 0.172$\pm$0.010 &  0.279$\pm$0.014 & 1.41$\pm$0.04 & 5.06 \\
$\begin{array}{l} {}^{26}\mathrm{Mg}(\textstyle{0}^+) \\ {}^{26}\mathrm{Si}(\textstyle{0}^+) \end{array} \Big\} \rightarrow  {}^{26} \mathrm{Al}(1^+\,1.06\,\mathrm{MeV})$  & 3.550(11) & 1.063$\pm$0.027 & 1.03$\pm$0.13 & 1.031$\pm$0.013 & 1.05$\pm$0.06  & 1.20$\pm$0.08 & 1.14\\
~${}^{32}$S$(\textstyle{0}^+$) $\leftrightarrow \,^{32}$Cl$(\textstyle{1}^+$)  & 6.74(18) & 0.0021$\pm$0.0009 & $6.9^{+7.3}_{-4.1}$   & 0.046$\pm$0.010  & 0.116$\pm$0.043  & 0.99$\pm$0.05 &  8.6 \\
~${}^{39}$K$(\textstyle{3 \over 2}^+$) $\leftrightarrow \,^{39}$Ca$(\textstyle{3 \over 2}^+$)  & 3.6326(10) & 1.060$\pm$0.008 & 1.41$\pm$0.17  & 1.030$\pm$0.004  & 1.22$\pm$0.07 &$3.1\pm 0.3$ & 2.5 \\
\hline
\hline
 \end{tabular}%
 \end{adjustbox}
  }
\end{table*}
\noindent
Here we make two modifications in the compiled $\mathrm{B_{GT}^{(p,n)}}$ values.  The first is a reduction by a factor of $(1.251/1.276)^2$ to account
for the current value of $g_A$.  The second addresses the absence of experimental errors on the 
compiled \cite{Watson} BGT values.  In previous work
a value for $\delta$ was obtained by a simple fit, which weights all data points equally and precludes a realistic estimate of the uncertainty for the derived  $\delta$.
As discussed below, we now include several new transitions in our analysis where uncertainties are available.  For the transitions we retain from the tabulation of \cite{Watson},
we have estimated uncertainties using \cite{Anderson83}, which \cite{Watson} references for experimental details.
The uncertainties tabulated there include the efficiency determination ($\pm$8\%), beam normalization ($\pm$5\%),
neutron attenuation ($\pm$5\%), counting statistics ($\pm$3\%), and background subtractions ($\pm$5\%).  Combining these in quadrature yields an estimated $\pm$12\%
uncertainty, which we adopt for all of the stronger transitions in Table \ref{tab:data}.  (Ref. \cite{Anderson83} also includes a correction for target water absorption, but that correction
addresses an issue specific to one target.)   The uncertainty inherent in (p,n) mappings of B$_\mathrm{GT}$ strength has been frequently discussed, with most
estimates in range of 10-20\% \cite{Zegers}.  Our choice of 12\% is consistent with this range.\\
\begin{figure*}[ht]   
\centering
\includegraphics[scale=0.39]{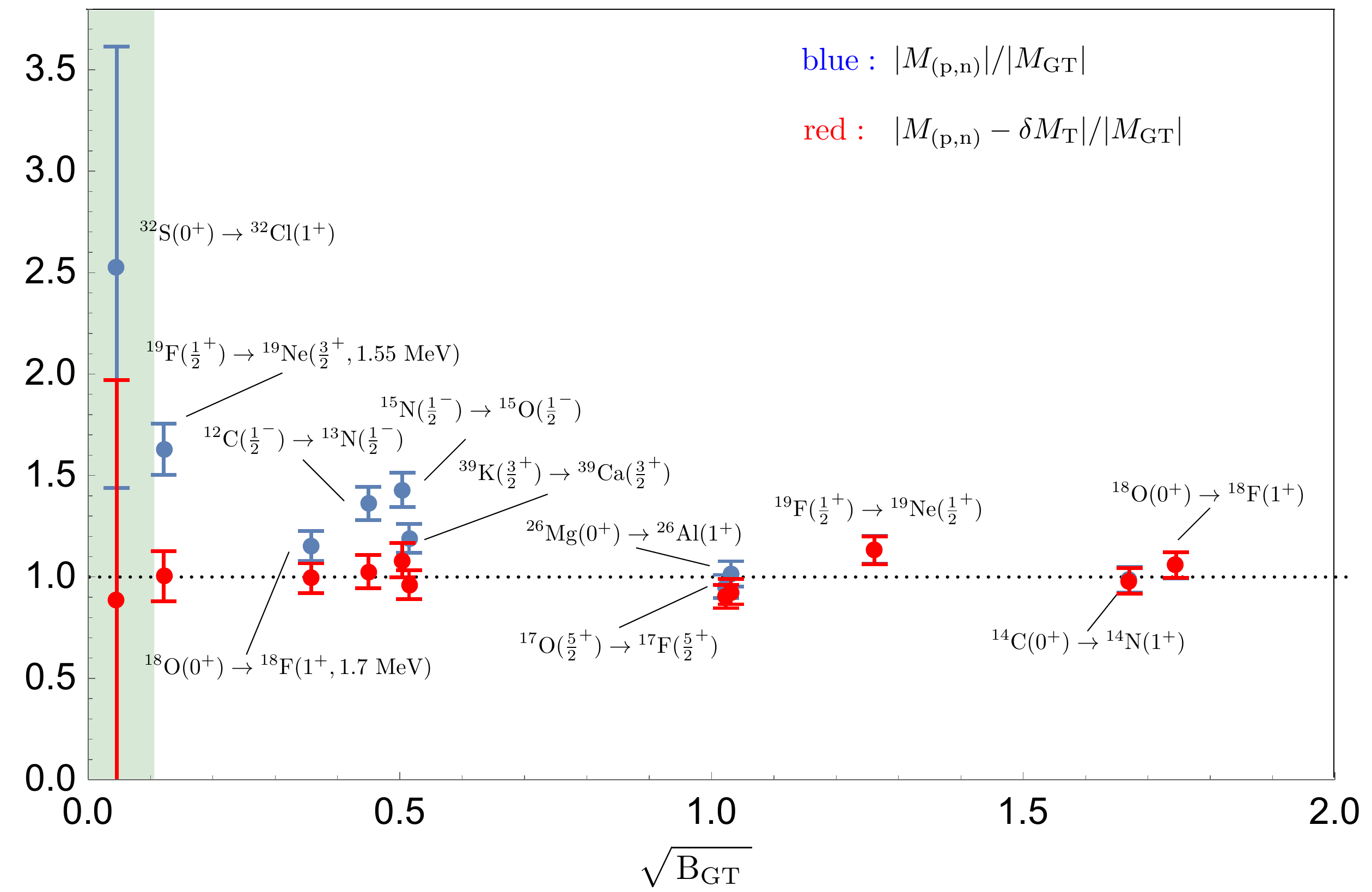}
\caption{In blue: correspondence between the (p,n) amplitude $|M^\mathrm{(p,n)}|$ and the beta decay amplitude $|M_\mathrm{GT}|$
is excellent when B$_\mathrm{GT}$ is strong, but deteriorates for weaker B$_\mathrm{GT}$.  In red: the agreement is restored with the inclusion of $|M_\mathrm{T}|$.
The two excited states that contribute to the BEST cross section have weak transition strengths that would place them in the shaded region,
where large tensor contributions would be anticipated.}
\label{fig2:BGT}
\end{figure*}
~\\
~\\
Two additional transitions used in \cite{HH1,HH2}, $^{32}$S$(0^+)\leftrightarrow\,^{32}$Cl($1^+$) and $^{39}$K(${3 \over 2}^+) \leftrightarrow\,^{39}$Cl(${1 \over 2}^+)$,
are candidate $\ell$-forbidden M1 transitions sensitive to the tensor amplitude.
One expects both to be dominated by the transition density $2s_{1/2} \leftrightarrow 1d_{3/2}$.
The $^{39}$K $M^\mathrm{(p,n)}$ given in \cite{HH1} was extracted from raw (p,n) cross sections as an order-of-magnitude estimate: there is no experimentally extracted B$_\mathrm{GT}^\mathrm{(p,n)}$ value.  No meaningful error
can be assigned to this very weak transition, so it is not included here.   The $^{32}$S transition was reconsidered in \cite{Grewe}, where a large uncertainty was assigned, which we adopt for our
analysis.  The impact of this transition on the current analysis is greatly diminished by the size of that uncertainty.\\
~\\
We also include two $2s1d$-shell transitions not considered in earlier analyses, one a recent result for $^{26}$Mg($0^+) \rightarrow\,^{26}$Al$(1^+,1.70\,$MeV), obtained
from $(^3$He,t).  While this result was normalized to the $^{26}$Al ground state $\beta$ decay rate, the unusually weak tensor contribution we predict for the normalizing transition ($\lesssim$ 1\%) allows
us to accept this result.  The second is $^{19}$F(${1 \over 2} ^+) \rightarrow\,^{19}$Ne$(\textstyle{3 \over 2}^+,1.55\,$MeV), a transition that is strongly $\ell$-forbidden, according to the
shell model, and thus potentially quite sensitive to $\hat{O}_\mathrm{T}^{J=1}$.\\
~\\
The data displayed in Table \ref{tab:data} include $\beta$ decay log(ft) values taken from the ENSDF data files \cite{ENSDF}, from which we determine $M_\mathrm{GT}$ and its uncertainty,
and calculations of $M_\mathrm{T}$ --- one needs both the magnitude and sign of this quantity relative to $M_\mathrm{GT}$.  In nine of the tabulated cases the shell-model calculations we perform (see below)
yield a positive relative sign, and thus a positive $\delta$ to account
for the observed enhancement in $|M^\mathrm{(p,n)}|/|M_\mathrm{GT}|$.  There are two exceptions:  The three shell model calculations we performed for $^{18}$O($0^+) \rightarrow\,^{18}$F$(1^+)$ all predict
a negative $M_\mathrm{T}$, but with a magnitude so small that it has no impact on our study.  For $^{19}$F($\textstyle{1 \over 2} ^+) \rightarrow\,^{19}$Ne$(\textstyle{3 \over 2}^+,1.55\,$MeV), the shell model
calculations disagree on the sign, and all underestimate the already quite suppressed $M_\mathrm{GT}$ derived from experiment.  We have assumed constructive interference as this is indicated by experiment and is consistent with
the calculation that best reproduces the known value of $|M_\mathrm{GT}|$ (the USDA shell model result described below).~\\
 ~\\
Our shell-model calculations of $M_\mathrm{T}$ were performed with the Cohen and Kurath \cite{CK} interaction in the $1p$ shell and the Brown-Wildenthal \cite{Wildenthal} and
USDA/USDB \cite{USD} interactions in the $2s1d$ shell.  The availability of three $2s1d$ interactions that each do well in reproducing $2s1d$ 
spectroscopy provides an opportunity to assess theory uncertainties.  The $2s1d$-shell values in the table are the means and standard deviations  of the three calculations.  There is excellent consistency, typically at the level of 10\%.  Even by eye, there is
clearly a strong correlation between the cases in Table \ref{tab:data} where $|M_\mathrm{T}|/|M_\mathrm{GT}|$ is large and those where
the experimental ratio $|\mathrm{B}_\mathrm{GT}^\mathrm{(p,n)}|/|\mathrm{B}_\mathrm{GT}|$ is significantly above 1. 
In the four cases where the $|\mathrm{B}_\mathrm{GT}^\mathrm{(p,n)}|/|\mathrm{B}_\mathrm{GT}|$ exceeds 1.4, $|M_\mathrm{T}|/|M_\mathrm{GT}|$ ranges from 2.5 to 8.6.\\
 ~\\
 We use these results to test whether the inclusion of the tensor operator in Eq. (\ref{eq:MT}) improves the agreement between $M^\mathrm{(p,n)}$ and $M_\mathrm{GT}$.  Evaluating the $\chi^2$ 
 per degree of freedom with and without $M_\mathrm{T}$ yields
 \begin{eqnarray}
 {1 \over 11} \sum_{i=1}^A { (M_\mathrm{GT}(i)-M^\mathrm{(p,n)}(i))^2 \over \sigma{(i)}^2 } &\approx& 9.4\nonumber \\
 {1 \over 10} \sum_{i=1}^A { (M_\mathrm{GT}(i)+\delta M_\mathrm{T}(i)-M^\mathrm{(p,n)}(i))^2 \over \sigma{(i)}^2 } &\approx& 1.0\nonumber \\
 & &
 \end{eqnarray}
where $\delta=0.076$, and $\sigma(i)$ is generated by combining uncertainties for $M^\mathrm{(p,n)}$ and $M_\mathrm{GT}$ in quadrature.  While a simple proportionality
between B$_\mathrm{GT}$ strength and (p,n) scattering for individual states is not supported by the data, the proportionality is restored with the introduction
of $M_\mathrm{T}$ to a level consistent with the statistical fluctuations of the data.
The variation around the minimum to achieve a unit change in the total $\chi^2$ yields the estimate $\delta = 0.076 \pm 0.008$ at $1\sigma$.~\\
~\\
The analysis can also be done by examining each target separately: this approach has advantages in understanding the
relationship of the current work to that of \cite{HH1,HH2}.  For each target $i$ we determine a probability distribution for $\delta(i)$ from the 
relation $M_\mathrm{GT}(i)+\delta(i) M_\mathrm{T}(i)-M^\mathrm{(p,n)}(i)=0$, treating the errors on $M_\mathrm{GT}(i)$, $M_\mathrm{T}(i)$, and $M^\mathrm{(p,n)}(i)$
as Gaussian with the uncertainties listed in Table \ref{tab:data}. The convolution was done by discretizing the probability distributions in bins, and independently by
Monte Carlo, and the results cross-checked to verify their numerical accuracy.
The theory errors on $M_\mathrm{T}$ (see Table \ref{tab:data}) were computed from the standard deviations of the SM results in the cases where multiple effective interactions were explored; in the five cases where only a single shell-model
calculation was done, we assigned an uncertainty of 10\%, a value typical of the other cases. (In the simple $\chi^2$ fit described previously, we used the best values for the $M_\mathrm{T}$, neglecting the uncertainties.) \\
~\\
The resulting probability distributions for the $\delta(i)$, while not exactly Gaussian, turn out to be nearly so in all cases.  
The equivalent Gaussian means and standard deviations are given in Table \ref{tab:stat}.  These results can then be combined to form the overall uncertainty-weighted mean and standard deviation, $\delta = 0.074 \pm 0.008~(1 \sigma)$,
a result nearly identical to that obtained more simply from the $\chi^2$.  This is our final result for $\delta$.\\
~\\
In this fit, the result is dominated by four transitions, from $^{13}$C,  $^{15}$N, $^{39}$K, and $^{19}$F to the excited state, which are four of the five cases where $M_\mathrm{T}$ exceeds $M_\mathrm{GT}$ 
by factors $\gtrsim$ 2.5.  If only these four transitions are retained, one obtains $\delta = 0.078 \pm 0.009$.   The other seven constraints have a minimal effect, shifting the mean by $\approx 5\%$ and improving the precision
by only $\approx 6\%$. This reflects the fact that in computing the weighted mean and uncertainty, the contributions of these seven are diluted by their low weights, $w_i = 1/\sigma_i^2$,
for the $\sigma_i$ of Table \ref{tab:stat}. Thus it is somewhat fortuitous that earlier work \cite{HH1,HH2} in which central values were fit, thereby weighting each target equally, gave results consistent with the range 
determined here.  (The results from \cite{HH1,HH2} are $\delta=0.096$ and 0.069 for the $1p$ and $2s1d$ shells, respectively.)  \\
~\\
These results and their relevance to the $^{71}$Ga($\nu_e,e^-)^{71}$Ge cross section are apparent from Fig. \ref{fig2:BGT}.  The agreement
between the (p,n) amplitude $|M^\mathrm{(p,n)}|$ and $|M_\mathrm{GT}|$, which is excellent for transitions with strong B$_\mathrm{GT}$ values,
systematically deteriorates as B$_\mathrm{GT}$ is reduced.  But this deterioration is corrected by the inclusion of $M_\mathrm{T}$.
The shaded region at small B$_\mathrm{GT}$ is that relevant for the
two $^{71}$Ga excited-state transitions: based on the trends apparent from the figure, the interpretation of (p,n) data for these transitions would not be reliable unless the
effects of $M_\mathrm{T}$ are treated.\\
~\\
From Fig. \ref{fig2:BGT} one sees that a fixed $\delta$ brings the (p,n) results into accord with known Gamow-Teller strengths throughout the $1p$
and $2s1d$ shells.  The absence of any evident dependence on mass number justifies the use of the same $\delta$ in our $^{71}$Ga cross section work.  It
would be helpful to verify this assumption by extending the results of Fig. \ref{fig2:BGT} into the $2p_{3/2}1f_{5/2}2p_{1/2}1g_{9/2}$ shell.  Obstacles to
doing this successfully include the absence of an experimental compilation for heavier nuclei analogous to that of \cite{Watson}, fewer opportunities to exploit isospin mirror transitions
(which play a major role in the analysis presented here), and the theory challenge of evaluating the tensor amplitudes in systems with higher level densities and
consequently more delicate level mixing.

\begin{table}
 \caption{The Gaussian means $\bar{\delta}_i$ and standard deviations $\sigma_i$ for the distributions $\delta_i$ obtained from
 each of the reactions.  The weighted combination of these results yields $\delta = 0.074 \pm 0.008~(1\sigma)$.    \label{tab:stat} }
\centering
{\renewcommand{\arraystretch}{1.6}
\begin{ruledtabular}
\begin{tabular}{lcc}
 Reaction &  \multicolumn{1}{c}{$\bar{\delta}_i$} & \multicolumn{1}{c}{$\sigma_i$}\\
\hline
~${}^{13}$C$(\textstyle{1 \over 2}^-$) $\leftrightarrow \, {}^{13} $N$(\textstyle{1 \over 2}^-$) & 0.082 & 0.020 \\
$\begin{array}{l}  {}^{14}\mathrm{C}(\textstyle{0}^+) \\   {}^{14}\mathrm{O}(\textstyle{0}^+) \end{array} \Big\} \rightarrow  {}^{14}\mathrm{N}(1^+\,3.95\,\mathrm{MeV})$  & -0.24 & 1.21 \\
~${}^{15}$N$(\textstyle{1 \over 2}^-$) $\leftrightarrow \,^{15}$O$(\textstyle{1 \over 2}^-$) & 0.093 & 0.021  \\
~${}^{17}$O$(\textstyle{5 \over 2}^+$) $\leftrightarrow \,^{17}$F$(\textstyle{5 \over 2}^+$)  & -0.070 & 0.085 \\
~${}^{18}$O$(\textstyle{0}^+$) $\leftrightarrow \,^{18}$F$(\textstyle{1}^+$)  & -1.59 & 2.48 \\[.1cm]
$\begin{array}{l}  {}^{18}\mathrm{O}(\textstyle{0}^+) \\  {}^{18}\mathrm{Ne}(\textstyle{0}^+)  \end{array} \Big\} \rightarrow {}^{18}\mathrm{F} (1^+\,1.70\,\mathrm{MeV})$ & 0.063 & 0.040 \\
~${}^{19}$F$(\textstyle{1 \over 2}^+$) $\leftrightarrow \,^{19}$Ne$(\textstyle{1 \over 2}^+$)  & 2.69 & 1.80 \\
~${}^{19}$F$(\textstyle{1 \over 2}^+$) $\leftrightarrow \,^{19}$Ne$(\textstyle{3 \over 2}^+$~1.55 MeV)  & 0.076 & 0.012 \\
$\begin{array}{l} {}^{26}\mathrm{Mg}(\textstyle{0}^+) \\ {}^{26}\mathrm{Si}(\textstyle{0}^+) \end{array} \Big\} \rightarrow  {}^{26} \mathrm{Al}(1^+\,1.06\,\mathrm{MeV})$  & 0.015 & 0.051 \\
~${}^{32}$S$(\textstyle{0}^+$) $\leftrightarrow \,^{32}$Cl$(\textstyle{1}^+$)  & 0.070  & 0.045  \\[.1cm]
~${}^{39}$K$(\textstyle{3 \over 2}^+$) $\leftrightarrow \,^{39}$Ca$(\textstyle{3 \over 2}^+$)  & 0.061 & 0.023 \\
 \end{tabular}
 \end{ruledtabular}
 }
\end{table}

\begin{table*}
\centering
 \caption{$^{71}$Ga SM transitions matrix elements evaluated for each of three interactions.  For transitions to the two excited states in $^{71}$Ge,
 the predicted values and $1 \sigma$ ranges for $M_\mathrm{T}$ are used in the extraction of  $|M_\mathrm{GT}|$ from forward-angle (p,n) measurements.  We 
 include the SM values for $M_\mathrm{GT}$, denoted  $M_\mathrm{GT}^\mathrm{SM}$, to illustrate that SM GT transition strengths vary considerably for
 weak transitions, even when highly tuned effective interactions are employed.  For this reason $M_\mathrm{GT}$ is extracted from experiment -- not from theory.
 Theory is used only in estimating $M_\mathrm{T}$, a correction in the (p,n) analysis that generically enters at the level of $\sim$ 8\%.  This strategy dilutes the
 impact of nuclear structure uncertainties.
  \label{tab:Ga} }
  \begin{ruledtabular}
\begin{tabular}{lcddc}
& & &  &  \\[-.2cm]
 Transition  &  Interaction &  \multicolumn{1}{c}{~$M_\mathrm{GT}^\mathrm{SM}$}  &\multicolumn{1}{c}{~$M_\mathrm{T}$}    & ~~~~$\overline{M}_\mathrm{T}$~~~~   \\[.1cm]
  \hline
 & & & &   \\[-.25cm]
 $\textstyle{3 \over 2}^- \rightarrow \textstyle{1 \over 2}^-$(gs) & JUN45 & 0.791 &-0.516 &    \\[.1cm]
 & GCN2850 & 0.361 & -0.320   & -0.37$\pm$0.13  \\[.1cm]
  & jj44b & 0.290  &-0.283 &   \\[.2cm]
  $\textstyle{3 \over 2}^- \rightarrow \textstyle{5 \over 2}^-$(175 keV) & JUN45 & 0.145 &-0.764  &    \\[.1cm]
 & GCN2850 & 0.159 & -0.410 &-0.50$\pm$0.24   \\[.1cm]
  & jj44b & 0.264 &-0.311 &   \\[.2cm]
  $\textstyle{3 \over 2}^- \rightarrow \textstyle{3 \over 2}^-$(500 keV) & JUN45 & 0.096 & 0.062  &    \\[.1cm]
 & GCN2850 & 0.196 & -0.178   &-0.088$\pm$0.13  \\[.1cm]
  & jj44b & 0.505 &-0.148 &   \\[.1cm]
 \end{tabular}
 \end{ruledtabular}
\end{table*}

\section{Recommended excited-state B$_\mathrm{{\bf GT}}$ values}
With $\delta = 0.074 \pm 0.008~(1 \sigma)$ determined,  we can now extract from (p,n) measurements best values and estimated uncertainties for $M_\mathrm{GT}$
for the two excited-state contributions to $^{71}$Ga$(\nu_e,e^-)^{71}$Ge, using
\[            M^\mathrm{(p,n)} = M_\mathrm{GT} + \delta M_\mathrm{T}. \]
This requires us to compute the magnitude and relative sign of $M_\mathrm{T}$.
The theory task is more challenging than that of the previous section because
the effective interactions available for the relevant shell-model space, $2p_{3/2}1f_{5/2}2p_{1/2}1g_{9/2}$, are known to be less 
successful in their spectroscopic predictions.  These nuclear physics uncertainties should be reflected in the range of the predicted $M_\mathrm{T}$.\\
~\\
As was done in the $2s1d$ shell, calculations were performed with three well-tested interactions,  GCN2850 \cite{GCN2850}, jj44b \cite{jj44b}, and JUN45 \cite{JUN45},
including all $m$-scheme Slater determinants that can be formed in the valence space ($\approx10^8$ basis states).
The results are shown in Table \ref{tab:Ga}.  There is reasonable agreement among these calculations on magnitudes and signs, and as before,
we combine the results to obtain best values and $1\sigma$ ranges.  The
combined results are denoted $\overline{M}_\mathrm{GT}$ and $\overline{M}_\mathrm{T}$.
The results for $\overline{M}_\mathrm{T}$ --- its magnitude and sign relative to $M_\mathrm{GT}$ ---
are needed in the analysis below.   An immediate cross check on the nuclear structure comes from the known B$_\mathrm{GT}$ value derived from the $^{71}$Ge electron capture rate, $B_\mathrm{GT}=0.0864\pm0.0013~(2 \sigma)$, 
from which one finds $M_\mathrm{GT} = 0.588 \pm 0.002~(1\sigma)$, in good agreement with the shell-model result, $\overline{M}_\mathrm{GT} = 0.48 \pm 0.27~(1\sigma)$.
After we extract the ${M}_\mathrm{GT}$'s from the (p,n) results, we will be able to make similar comparisons for the excited states.  Note that
the shell model indicates largely destructive interference between the $M_\mathrm{GT}$ and $M_\mathrm{T}$ amplitudes for the
three transitions of interest.\\
~\\
{\it Analysis for Krofcheck et al.:} The forward-angle (p,n) scattering results of Krofcheck et al. \cite{Krofcheck} for exciting the $^{71}$Ge ground-state ($\textstyle{1 \over 2}^-$), 175 keV  ($\textstyle{5 \over 2}^-$), and 500 keV ($\textstyle{3 \over 2}^-$) levels yield B$_\mathrm{GT}^\mathrm{(p,n)}$ = 0.089$\pm$0.007, $<0.005$, and 0.011$\pm$0.002, respectively.  These results were normalized to the analog transition using 
an energy-dependent
coefficient relating B$_\mathrm{GT}^\mathrm{(p,n)}$ to cross sections.  As we are concerned with states in a narrow energy band, we can avoid any
issues with this choice by forming the ratios
\begin{eqnarray}
{ \mathrm{B_{GT}^{(p,n)}}(\textstyle{5 \over 2}^-) \over \mathrm{B_{GT}^{(p,n)}}(\mathrm{gs})}&\equiv&{|M_\mathrm{GT}(\textstyle{5 \over 2}^-) + \delta \overline{M}_\mathrm{T}(\textstyle{5 \over 2}^-)|^2 \over |M_\mathrm{GT}(\mathrm{gs}) + \delta \overline{M}_\mathrm{T}(\mathrm{gs}) |^2} \\
\nonumber &<& 0.06 ~~~(68\%~\mathrm{C.L.}) \nonumber \\[0.2cm]  
{ \mathrm{B_{GT}^{(p,n)}}(\textstyle{3 \over 2}^-) \over \mathrm{B_{GT}^{(p,n)}}(\mathrm{gs})} &\equiv&{|M_\mathrm{GT}(\textstyle{3 \over 2}^-) + \delta \overline{M}_\mathrm{T}(\textstyle{3 \over 2}^-)|^2 \over |M_\mathrm{GT}(\mathrm{gs}) + \delta \overline{M}_\mathrm{T}(\mathrm{gs})|^2} \nonumber \\
&=& 0.124 \pm 0.024~~~(68\%~\mathrm{C.L.}) 
\label{eq:BGTvals}
\end{eqnarray}
~\\
We first consider the transition to the $\textstyle{3 \over 2}^-$ state.  
Because the electron capture $M_\mathrm{GT}$ is
so precisely known, we use the central value in the analysis below.  
We insert the values for the two tensor matrix elements $M_\mathrm{T}$ and $\delta$, including their uncertainties.  We take the (dominant) sign of $M_\mathrm{T}$ relative to $M_\mathrm{GT}$ from theory.  We assume
all distributions are normal (Gaussian), described by the specified standard deviations, then compute the associated distribution for the needed B$_\mathrm{GT}$ ratio, which
we find is also well represented by a normal distribution.  This yields
\begin{eqnarray}
{ \mathrm{B_{GT}}(\textstyle{3 \over 2}^-)  \over \mathrm{B_{GT}}(\mathrm{gs}) }& \equiv& {|M_\mathrm{GT}(\textstyle{3 \over 2}^-)|^2 \over |M_\mathrm{GT}(\mathrm{gs}) |^2} \nonumber \\
&=& 0.121 \pm 0.026~~~(68\%~\mathrm{C.L.})~.
\label{eq:500}
\end{eqnarray}
The tensor corrections have little impact, shifting the numerator and denominator similarly, by 3.2\% and 4.7\% respectively, with these shifts largely canceling when the ratio is formed.
The central value obtained for $|M_\mathrm{GT}(\textstyle{3 \over 2}^-)| \, \approx \, 0.20$ is consistent with the
shell-model range in Table \ref{tab:Ga}, 0.27 $\pm$ 0.21.\\
~\\
In our shell-model calculations the density matrices for the transition to the 175 keV $\textstyle{5 \over 2}^-$ state are dominated by the
the $\ell$-forbidden amplitude $2p_{3/2} \rightarrow 1f_{5/2}$, which reaches single-particle strength in the case of the JUN45 calculation.  This is the reason for the strength of $M_\mathrm{T}$
and the weakness of $M_\mathrm{GT}$ in the shell-model studies of Table \ref{tab:Ga}, and consequently $\overline{M}_\mathrm{T}/\overline{M}_\mathrm{GT} \, \approx \, -2.6$.
The destructive interference allows for a larger $|M_\mathrm{GT}|$ than would be the case if the tensor contributions to (p,n) scattering were ignored.
As was done for the 500 keV state, we take into account the uncertainties on the various quantities by integrating over the probability distributions of each
input variable, taking the ranges of the $\overline{M}_\mathrm{T}$ from the results of Table \ref{tab:Ga}.  
In this calculation we interpret the experimental bound given in the first of Eqs. (\ref{eq:BGTvals})  as a measurement of 0 
with a one-sided normal distribution described by $\sigma=0.06$.  We find
\begin{eqnarray}
|M_\mathrm{GT}(\textstyle{5 \over 2}^-)| &\lesssim& \left\{  \begin{array}{ll} 0.18 & ~68\% ~\mathrm{C.L.} \\[.07cm] 0.24 & ~95\% ~\mathrm{c.l} \end{array} \right. \nonumber \\ \displaystyle{ \mathrm{B_{GT}}(\textstyle{5 \over 2}^-)  \over \mathrm{B_{GT}}(\mathrm{gs}) } &\lesssim& \left\{ \begin{array}{ll}  0.089  &  ~68\%~\mathrm{C.L.} \\[.07cm]  0.160 & ~95\%~\mathrm{C.L.} \end{array} \right.  
\label{eq:175}
\end{eqnarray}
~\\
\begin{figure}[ht]   
\centering
\includegraphics[scale=0.68]{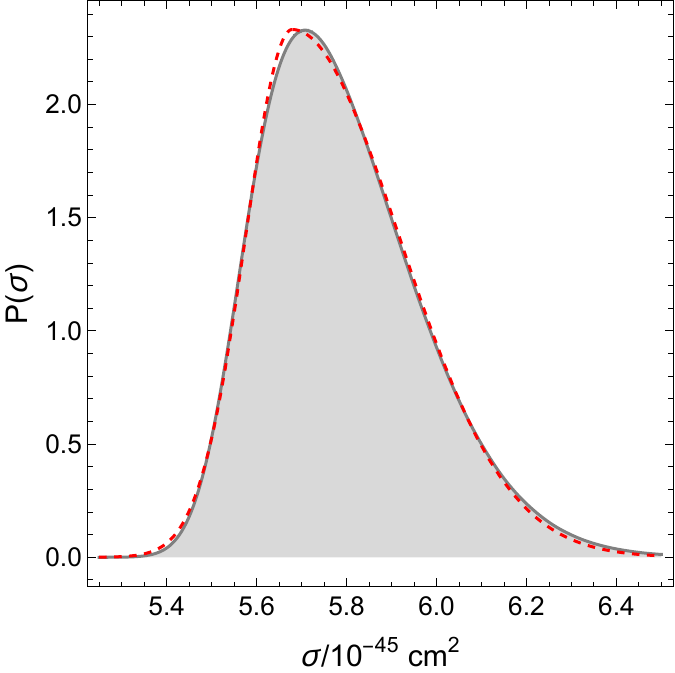}
\caption{Shaded region: numerically generated probability distribution for the $^{51}$Cr cross section with excited-state
contributions extracted from forward-angle (p,n) cross sections of \cite{Krofcheck} (see text).
Dashed line: the analytic split-normal fit to these data.}
\label{fig:splitn}
\end{figure}
{\it Analysis for Frekers et al.:} We repeat the analysis for the Frekers et al. \cite{FrekersGa} data, 
as the use of the same effective operator for $(^3$He,t) has support from both theory and experiment \cite{Frekers}, 
while noting that a separate derivation of $\delta$ based on data like those of Table \ref{tab:data} has not been done for this reaction.  We again form the ratios
\begin{eqnarray}
{ \mathrm{B_{GT}^{(^3He,t)}}(\textstyle{5 \over 2}^-) \over \mathrm{B_{GT}^{(3He,t)}}(\mathrm{gs})}&\equiv&{|M_\mathrm{GT}(\textstyle{5 \over 2}^-) + \delta \overline{M}_\mathrm{T}(\textstyle{5 \over 2}^-)|^2 \over |M_\mathrm{GT}(\mathrm{gs}) + \delta \overline{M}_\mathrm{T}(\mathrm{gs}) |^2} \nonumber\\
 &=& 0.040 \pm 0.031 ~~~(68\%~\mathrm{C.L.}) \nonumber  \\[0.2cm]  
{ \mathrm{B_{GT}^{(^3He,t)}}(\textstyle{3 \over 2}^-) \over \mathrm{B_{GT}^{(3He,t)}}(\mathrm{gs})} &\equiv&{|M_\mathrm{GT}(\textstyle{3 \over 2}^-) + \delta \overline{M}_\mathrm{T}(\textstyle{3 \over 2}^-)|^2 \over |M_\mathrm{GT}(\mathrm{gs}) + \delta \overline{M}_\mathrm{T}(\mathrm{gs})|^2}\nonumber \\
&=& 0.207 \pm 0.019~~~(68\% ~\mathrm{C.L.})
\label{eq:BGTs}
\end{eqnarray}
~\\
The calculation for the 500 keV $\textstyle{3 \over 2}^-$ excited state state proceeds as before, 
taking into account the values and uncertainties for the two tensor matrix elements $M_\mathrm{T}$ and $\delta$, and taking the
relative signs of the $M_\mathrm{T}$ from theory.    This yields
\begin{equation}
{ \mathrm{B_{GT}}(\textstyle{3 \over 2}^-)  \over \mathrm{B_{GT}}(\mathrm{gs}) } = 0.198 \pm 0.024 ~~~(68\%~\mathrm{C.L.})~.
\label{eq:500F}
\end{equation}
Thus the effects of $M_\mathrm{T}$ are modest, for the reasons mentioned above.  The central value $M_\mathrm{GT}(\textstyle{3 \over 2}^-) \, \approx \, 0.26$ is consistent with the theory range of Table \ref{tab:Ga},  $0.27 \, \pm \, 0.21$.  \\
~\\
A similar analysis for the Frekers result for the transition to the 175 keV state yields
\begin{eqnarray}
   \left. \begin{array}{r} 0.11 \\[.07cm]  0.065 \end{array}  \right\}  &\lesssim& |M_\mathrm{GT}(\textstyle{5 \over 2}^-)| \lesssim \left\{  \begin{array}{ll} 19 & ~68\% ~\mathrm{C.L.} \\[.07cm] 0.23 & ~95\% ~\mathrm{c.l} \end{array} \right. \nonumber \\
   \displaystyle{ \mathrm{B_{GT}}(\textstyle{5 \over 2}^-)  \over \mathrm{B_{GT}}(\mathrm{gs}) } &=& 0.071 \pm 0.036 ~~~(68\%~\mathrm{C.L.}) .
   \label{eq:175F}
   \end{eqnarray} 
The central value for $\left|M_\mathrm{GT}\left(\frac{5}{2}^-\right)\right| \, \approx \, 0.15$ is consistent with the shell-model range, 0.19$\pm$0.07.  ~\\
\section{Total Cross Sections}
The  total neutrino cross sections for 
$^{51}$Cr and ${}^{37}$Ar can be expressed in the form \cite{HH1}
\begin{equation}
 \sigma = \sigma_\mathrm{gs} \left[ 1 + \xi(\textstyle{ 5 \over 2}^-)\, \displaystyle{ \mathrm{B_{GT}}(\textstyle{5 \over 2}^-)  \over \mathrm{B_{GT}}(\mathrm{gs}) }+ \xi(\textstyle{ 3 \over 2}^-)\,\displaystyle{ \mathrm{B_{GT}}(\textstyle{3 \over 2}^-)  \over \mathrm{B_{GT}}(\mathrm{gs}) } \right]
 \end{equation}
where the phase-space coefficients, computed from the results of Table \ref{tab:neutrino_capture}, are
\begin{eqnarray} 
^{51}\mathrm{Cr}:~~ \xi(\textstyle{ 5 \over 2}^-)&=&0.669~~~\xi(\textstyle{ 3 \over 2}^-)=0.220\nonumber \\
^{37}\mathrm{Ar}:~~ \xi(\textstyle{ 5 \over 2}^-)&=&0.696~~~\xi(\textstyle{ 3 \over 2}^-)=0.264
\end{eqnarray}
(For earlier calculations of these coefficients see \cite{HH1,Bahcall97}.)\\
~\\
Combining the excited-state B$_\mathrm{GT}$ ratios derived from the data of Krofcheck et al. \cite{Krofcheck},  Eqs. (\ref{eq:500}) and (\ref{eq:175}), with
the ground-state result of Eq. (\ref{eq:gs}) yields
\begin{eqnarray}
\sigma(^{51}\mathrm{Cr}) &=& \left\{ \begin{array}{ll} 5.71^{+0.27}_{-0.10}  &   68\% ~\mathrm{C.L.} \\[.2cm]    5.71^{+0.51}_{-0.23}  &   95\%~\mathrm{C.L.}  \end{array} \right\}   \times 10^{-45}  
 \,\mathrm{cm}^2\nonumber \\
 & & \\
 \sigma(^{37}\mathrm{Ar}) &=&  \left\{ \begin{array}{ll} 6.88^{+0.34}_{-0.13}  &   68\%~\mathrm{C.L.} \\[.2cm]    6.88^{+0.63}_{-0.28}  &   95\%~\mathrm{C.L.}  \end{array} \right\}  \times 10^{-45} \,\mathrm{cm}^2 \nonumber
 \label{eq:K}
\end{eqnarray}
The probability distributions for the cross sections were computed numerically by folding the ground-state and excited-state probabilities.
The central values correspond to the most probable cross section and the ranges contain the 68\% and 95\% fractions of the most probable results.
The excited-state contributions increase the cross sections by $\approx$ 6.0\% and 6.6\% for $^{51}$Cr and $^{37}$Ar, respectively.\\
~\\
Repeating this calculation using the excited-state B$_\mathrm{GT}$ ratios derived from the data of Frekers et al. \cite{FrekersGa},  Eqs. (\ref{eq:500F}) and (\ref{eq:175F}),  yields

\begin{eqnarray}
 \sigma(^{51}\mathrm{Cr}) &=& \left\{ \begin{array}{ll} 5.85^{+0.19}_{-0.13}  &   68\%~\mathrm{C.L.} \\[.2cm]    5.85^{+0.36}_{-0.26}  &   95\%~\mathrm{C.L.}  \end{array} \right\}   \times 10^{-45}  
 \,\mathrm{cm}^2\nonumber \\
 & & \\
 \sigma(^{37}\mathrm{Ar}) &=&  \left\{ \begin{array}{ll} 7.01^{+0.22}_{-0.16}  &   68\%~\mathrm{C.L.} \\[.2cm]    7.01^{+0.43}_{-0.31}  &   95\%~\mathrm{C.L.}  \end{array} \right\}  \times 10^{-45} \,\mathrm{cm}^2\nonumber
 \label{eq:F}
\end{eqnarray}
The excited-state contributions increase the cross sections by $\approx$ 8.6\% for both $^{51}$Cr and $^{37}$Ar sources.  The results of Eqs. (\ref{eq:K}) and (\ref{eq:F}) agree at $1 \sigma$, but in our
view should not be combined because they depend on input strengths for the transition to the $\textstyle{3 \over 2}^-$ state that differ by significantly more.  \\
~\\
The numerically evaluated cross section distributions can be accurately described as split-normal probability distributions
\begin{eqnarray}
P(\sigma)\approx \sqrt{2 \over \pi}&~ \displaystyle{1 \over \sigma_1 + \sigma_2} ~\bigg[ \theta(\bar{\sigma} -\sigma) e^{-(\sigma-\bar{\sigma})^2/2\sigma_1^2}\nonumber \\
&+\theta(\sigma-\bar{\sigma}) e^{-(\sigma-\bar{\sigma})^2/2 \sigma_2^2} \bigg],
\end{eqnarray}
where $\sigma$ is dimensionless, in units of $10^{-45}$ cm$^2$.  The best fits are obtained
by tuning the parameters given above to optimize the overall fit. Figure \ref{fig:splitn}, for the case of a $^{51}$Cr source and
Krofcheck et al. \cite{Krofcheck} excited state contributions, illustrates the quality of the fit.  These split-normal distributions will enable users to adapt our results for any desired confidence level.  
Table \ref{tab:split} gives the numerical values for the fit parameters.

\begin{table}
\centering
\caption{The split normal parameterization of our cross section results. \label{tab:split} }
  \begin{ruledtabular}
\begin{tabular}{lcccc}
& & &   & \\[-.2cm]
 Source  &  Excited States & $\bar{\sigma}$ & $\bar{\sigma}_1$ & $\bar{\sigma}_2$  \\[.1cm]
  \hline
 & & &  &  \\[-.25cm]
$^{51}$Cr & \cite{Krofcheck} & 5.68 & 0.10 & 0.24   \\[.1cm]
$^{51}$Cr & \cite{FrekersGa} & 5.83 & 0.12 & 0.17   \\[.1cm]
$^{37}$Ar & \cite{Krofcheck} & 6.84 & 0.13 & 0.30   \\[.1cm]
$^{37}$Ar & \cite{FrekersGa} & 6.99 & 0.15 & 0.21   \\[.1cm]
 \end{tabular}
 \end{ruledtabular}
\end{table}
~\\
\noindent
{\it Comparisons to past work:}  In Table \ref{tab:CrossSectionEstimates} we compare our cross section result to those
obtained by other authors in past years.  We briefly comment on the different approaches taken,
summarizing a more complete discussion that appears in \cite{PPNP}.
\begin{enumerate}[leftmargin=12pt]
 \item  Bahcall (1997) \cite{Bahcall97}:  This work included, in its estimate of $\sigma_{gs}$, overlap and exchange atomic effects, and used the then
 prevailing value of $Q_\mathrm{EC}$ = 232.69 $\pm$ 0.15 keV.   Excited state B$_\mathrm{GT}$ values were taken from the (p,n) values of \cite{Krofcheck}.
\item Haxton (1998) \cite{HH2}:  This work pointed out the need to include the tensor interaction when using (p,n) data, and estimated $M_\mathrm{T}$ from
a truncated SM calculation, as spaces of dimension $\sim 10^8$ could not be treated at the time.  Because this limited the included correlations, the SM value for $M_\mathrm{T}$ was taken as an upper bound,
yielding a large uncertainty on the extracted excited-state contributions.
 \item Barinov {\it et al.} (2018) \cite{Barinov2018}:  This work used weak couplings updated to 2018 and a value for $Q_\mathrm{EC} =233.5 \pm 1.2$ keV
 obtained from a Penning trap measurement of the mass difference that was later superseded by the more accurate trapping result of \cite{Frekers2016}.  The excited-state GT strengths were extracted from
 the (p,n) data, without  tensor corrections.
 \item  Kostensalo {\it et al.} (2019) \cite{Kostensalo}:  
 The cross section is taken from SM calculations using the JUN45 interaction, which among the interactions studied here predicts the smallest excited-state GT strengths.
 From \cite{Bahcall97} onward investigators have concluded that cross section estimates must be taken from experiment, with theory employed only for corrections (as has been done here):
 SM wave functions are soft projections (at best) of the true wave function, so lack many of the correlations important in evaluating the interfering amplitudes often
 responsible for weak transitions.
 
 The JUN45 B$_\mathrm{GT}$ values for $^{71}$Ga($\nu$,e)$^{71}$Ge to the $\textstyle{3 \over 2}^-$ and $\textstyle{5 \over 2}^-$ excited states
 are 2.3$\times10^{-3}$ and 5.2$\times 10^{-3}$, respectively.   We can test the predictive
 power of JUN45 using transitions of similar but known strengths in closely related nuclei.
 The B$_\mathrm{GT}$ values for $^{71}$As(EC)$^{71}$Ge to the $\textstyle{3 \over 2}^-$ and $\textstyle{5 \over 2}^-$ states of interest are known:
 $^{71}$As differs from $^{71}$Ga only by the conversion of a neutron pair to a proton pair.  There are similar testing opportunities 
 using $^{69}$Ge(EC)$^{69}$Ga, which involves parent and daughter nuclei differing from $^{71}$Ga and $^{71}$Ge only by the removal of a neutron pair. The results are given
 in Table \ref{tab:JUN45} and show large discrepancies between predicted and measured B$_\mathrm{GT}$ values, in two cases 
 exceeding an order of magnitude.  That is, the table is not encouraging.
 \item Semenov (2020) \cite{Semenov2020}:  This work follows \cite{Bahcall97} quite closely, treating the excited states as was done there, but utilizing updated weak couplings and 
 and the modern $Q_{EC}$ value of  \cite{Frekers2016}.
 \end{enumerate}
 ~\\
 In previous work, the determinations of $\sigma_{gs}$ have neglected a series of $\sim 0.5$\% effects 
 that we have addressed here, including Coulomb corrections computed from realistic nuclear densities consistent
 with the measured r.m.s. charge radius, weak magnetism corrections, and the difference in the radiative correction from
 bremsstrahlung to the EC and $(\nu,e)$ reactions.  Here we have addressed such corrections.
 
 Most past work has also taken excited-state
 contributions directly from forward-angle (p,n) reactions, assuming that a procedure calibrated for
 strong B$_\mathrm{GT}$ transitions and gross B$_\mathrm{GT}$ profiles could be applied to individual weak transitions.
 However, one expects the typical correction due to  
 $M_\mathrm{T}$ to be more important when the dominant amplitude with which it interferes, $M_\mathrm{GT}$,
 is suppressed.  This physics, apparent from Fig. \ref{fig2:BGT}, has been treated here with as
 much statistical rigor as possible, propagating input experimental and theoretical errors through to the
 extracted excited-state cross sections, to quantify their likelihoods.   This procedure is limited by the
 need to quantify the uncertainty on the correction $M_\mathrm{T}$, which must be taken from nuclear models.
 It is helpful that in the situation of most concern -- a weak $M_\mathrm{GT}$ interfering with a strong $M_\mathrm{T}$,
 thereby compensating in part for the small value of $\delta$ -- theory is needed only for the
 strong matrix element, as the sum is constrained by experiment.   The SM has a better track record in such cases.
 We discussed a common example, an $\ell$-forbidden M1 transition, where a weak $M_\mathrm{GT}$ and a strong $M_\mathrm{T}$ would arise.
 Here we have used the variation among SM predictions of $M_\mathrm{T}$ to define an uncertainty, with the understanding  
 that there could be additional hidden uncertainties, reflecting common assumptions of the SM affecting all calculations.

\begin{table}
\centering
\caption{A summary of the published neutrino reaction cross section estimates for $^{71}$Ga($\nu_e,e^-$)$^{71}$Ge in units of $10^{-45}$cm$^2$.  All results are given at 68\% C.L. 
\label{tab:CrossSectionEstimates}}
  \begin{ruledtabular}
\begin{tabular}{lccc}
& & &    \\[-.2cm]
     Author 	& Year	&  $\sigma(^{51}${Cr})	& $\sigma(^{37}${Ar}) 		 \\[.1cm]
  \hline
 & & &    \\[-.2cm]
Bahcall~\cite{Bahcall97}			&1997	&$5.81^{+0.21}_{-0.16}$ 	& $7.00^{+0.49}_{-0.21}$							\\
Haxton~\cite{HH2}				& 1998	&$6.39\pm0.68$ 					& 		--												\\
Barinov \textit{et al.}~\cite{Barinov2018}	 & 2018 	&$5.91\pm0.11$			& $7.14\pm0.15$			 		 				\\
Kostensalo \textit{et al.}~\cite{Kostensalo} & 2019	& $5.67\pm0.06$			& $6.80\pm0.08$											\\
Semenov~\cite{Semenov2020}		& 2020		& $5.94\pm0.12$			&$7.17\pm0.15$			 				\\
Present work    & 2023            & 5.71$^{+0.27}_{-0.10}$       &      6.88$^{+0.34}_{-0.13}$                               \\
 \end{tabular}
 \end{ruledtabular}
\end{table}
~\\
\begin{table}
\centering
\caption{Tests of JUN45 against neighboring transitions where log(ft) values are known from experiment. \label{tab:JUN45} }
  \begin{ruledtabular}
\begin{tabular}{lccc}
 & & &    \\[-.2cm]
 Transition  &   log(ft)  & B$_\mathrm{GT}^\mathrm{exp}$ & B$_\mathrm{GT}^\mathrm{JUN45}$    \\[.1cm]
  \hline
 & & &    \\[-.25cm]
${}^{71} \mathrm{As(EC)}^{71}\mathrm{Ge} (\textstyle{5 \over 2}^-,175 \, \mathrm{keV})$  & 5.85 & 5.3$\times10^{-3}$ & 6.9$\times10^{-3}$   \\[.1cm]
${}^{71} \mathrm{As(EC)}^{71}\mathrm{Ge} (\textstyle{3 \over 2}^-,500 \, \mathrm{keV})$ & 7.19 & 2.4$\times10^{-4}$ & 1.8$\times10^{-5}$   \\[.1cm]
${}^{69}\mathrm{Ge(EC)}^{69}\mathrm{Ga} (\textstyle{3 \over 2}^-, \mathrm{g.s.})$ & 6.49 & 1.2$\times10^{-3}$ & 3.4$\times10^{-5}$   \\[.1cm]
${}^{69}\mathrm{Ge(EC)}^{69}\mathrm{Ga} (\textstyle{5 \over 2}^-,574 \, \mathrm{keV})$ & 6.24 & 2.2$\times10^{-3}$ & 4.6$\times10^{-3}$   \\[.1cm]
 \end{tabular}
 \end{ruledtabular}
\end{table}
~\\
\section{Impact on the BEST and Gallium Anomalies}
The BEST $^{71}$Ge  production rates for the outer and inner volumes, obtained from the yields in the K and L peaks with a correction for the contribution
of the M peak \cite{BEST1,BEST2}, are
\begin{eqnarray}
R_\mathrm{out} &=& 55.6 \pm 2.7\, \mathrm{(stat)}^{+1.6}_{-1.5}\, \mathrm{(syst)}~\mathrm{atoms/d}  \nonumber \\
&=& 55.6 \pm 3.1 \mathrm{~atoms/d} \nonumber\\
R_\mathrm{in} &=& 54.9 \pm 2.5 \, \mathrm{(stat)}^{+1.6}_{-1.5} \, \mathrm{(syst)}~\mathrm{atoms/d} \nonumber \\
&=& 54.9^{+3.0}_{-2.9} \mathrm{~atoms/d}
\end{eqnarray}
where the statistical and systematic errors have been combined in quadrature to obtain a total error.  From the
neutrino source activity of 3.414$\pm$0.008 MCi and the cross section of Bahcall \cite{Bahcall97} used in \cite{BEST2}, one finds the predicted
production rates 
\begin{eqnarray}
R_\mathrm{out}^\mathrm{expected} &=& 72.6^{+2.6}_{-2.1} \mathrm{~atoms/d} \nonumber\\
R_\mathrm{in}^\mathrm{expected} &=& 69.4^{+2.5}_{-2.0} \mathrm{~atoms/d}
\end{eqnarray}
where again uncertainties have been combined in quadrature.  The ratios of the measured to
predicted production rates are 
\begin{eqnarray}
{ R_\mathrm{out} \over R_\mathrm{out}^\mathrm{expected}} &=& {55.6 \pm 3.1 \over 72.6^{+2.6}_{-2.1}}= 0.77 \pm 0.05  \nonumber\\
{ R_\mathrm{in} \over R_\mathrm{in}^\mathrm{expected}} &=&~ {54.9^{+3.0}_{-2.9}  \over 69.4^{+2.5}_{-2.0}}~ = 0.79 \pm 0.05 
\end{eqnarray}
showing discrepancies of 4.7 and 4.2 standard deviations, respectively.  The ratio of the outer and inner rates
\begin{equation}
r={R_\mathrm{out} \over R_\mathrm{in} }={0.766 \pm 0.05 \over 0.791 \pm 0.05} = 0.97 \pm 0.08
\end{equation}
is consistent with unity.

The Bahcall cross section used above employed the (p,n) data of Krofcheck et al. \cite{Krofcheck} in estimating the excited-state contribution to the $^{71}$Ga cross section.
The analogous analysis presented here, updating the ground-state contribution and correcting for the tensor contribution to the (p,n) results of \cite{Krofcheck}, yields
\begin{eqnarray}
R_\mathrm{out}^\mathrm{expected} &=& 71.4^{+3.4}_{-1.3} \mathrm{~atoms/d} \nonumber\\
R_\mathrm{in}^\mathrm{expected} &=& 68.2^{+3.2}_{-1.2} \mathrm{~atoms/d}
\end{eqnarray}
The ratios of the measured to predicted production rates are 
\begin{eqnarray}
{ R_\mathrm{out} \over R_\mathrm{out}^\mathrm{expected}} &=& {55.6 \pm 3.1 \over 71.4^{+3.4}_{-1.3}}= 0.78 \pm 0.05  \nonumber\\
{ R_\mathrm{in} \over R_\mathrm{in}^\mathrm{expected}} &=&~ {54.9^{+3.0}_{-2.9}  \over 68.2^{+3.2}_{-1.2}}~ = 0.80 \pm 0.05
\end{eqnarray}
As the cross section derived here is slightly reduced from that \cite{Bahcall97} used in the original BEST analysis \cite{Bahcall97},
the deviation of $R$ from 1 is also slightly reduced.

We have also constrained the $^{71}$Ge excited-state contribution using the ($^3$He,t) data of Frekers et al. \cite{FrekersGa}.
This yields
\begin{eqnarray}
R_\mathrm{out}^\mathrm{expected} &=& 73.1^{+2.4}_{-1.7} \mathrm{~atoms/d} \nonumber\\
R_\mathrm{in}^\mathrm{expected} &=& 69.9^{+2.3}_{-1.6} \mathrm{~atoms/d}
\end{eqnarray}
The ratios of the measured to predicted production rates are 
\begin{eqnarray}
{ R_\mathrm{out} \over R_\mathrm{out}^\mathrm{expected}} &=& {55.6 \pm 3.1 \over 73.1^{+2.4}_{-1.7}}= 0.76 \pm 0.05  \nonumber\\
{ R_\mathrm{in} \over R_\mathrm{in}^\mathrm{expected}} &=&~ {54.9^{+3.0}_{-2.9}  \over 69.9^{+2.3}_{-1.6}}~ = 0.79 \pm 0.05
\end{eqnarray}
reflecting the somewhat stronger transition to the $\textstyle{3 \over 2}^-$ extracted from the ($^3$He,t) data.

The combined result from all six Ga calibration experiments \cite{BEST2} for the ratio of the measured to expected rates are 0.82 $\pm$ 0.03,
using the cross section derived here and taking the excited-state data from \cite{Krofcheck}.  However, as discussed in \cite{PPNP}, when possible correlations among the measurements
are taken into account, this is revised to 0.81 $\pm$ 0.05.

\begin{figure*}[ht]   
\centering
\includegraphics[scale=0.32]{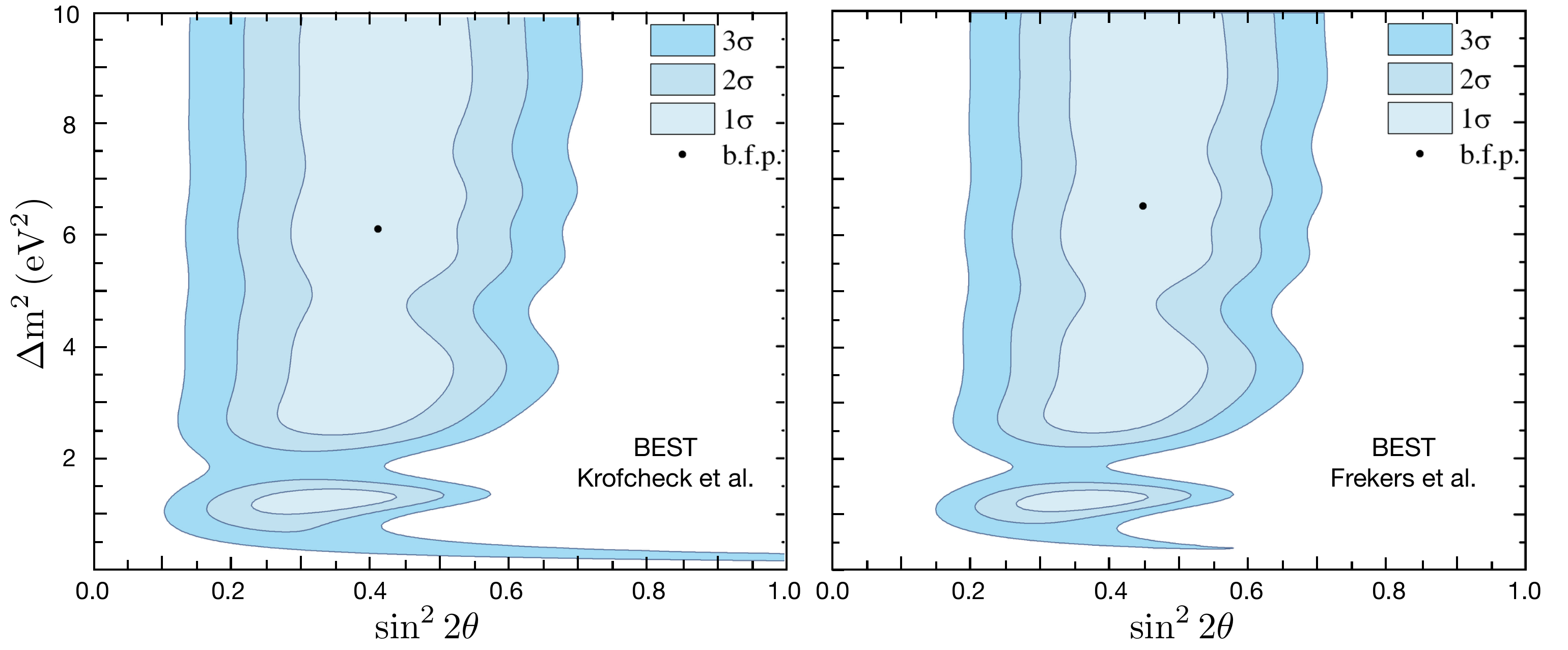}
\caption{The allowed regions for oscillations into a sterile state derived from the BEST inner and outer results.  
Left: the neutrino absorption cross section derived using
the (p,n) data of \cite{Krofcheck} to constrain excited-state contributions.  Right: results using the ($^3$He,t) data of \cite{FrekersGa}.   The best-fit points
are $\mathrm{sin}^2{2 \theta} =0.41$ and $\Delta \mathrm{m}^2=6.1$ eV$^2$ and $\mathrm{sin}^2{2 \theta} =0.45$ and $\Delta \mathrm{m}^2=6.5$ eV$^2$, respectively (see text).}
\label{fig:Fig4}
\end{figure*}

\begin{figure*}[ht]   
\centering
\includegraphics[scale=0.32]{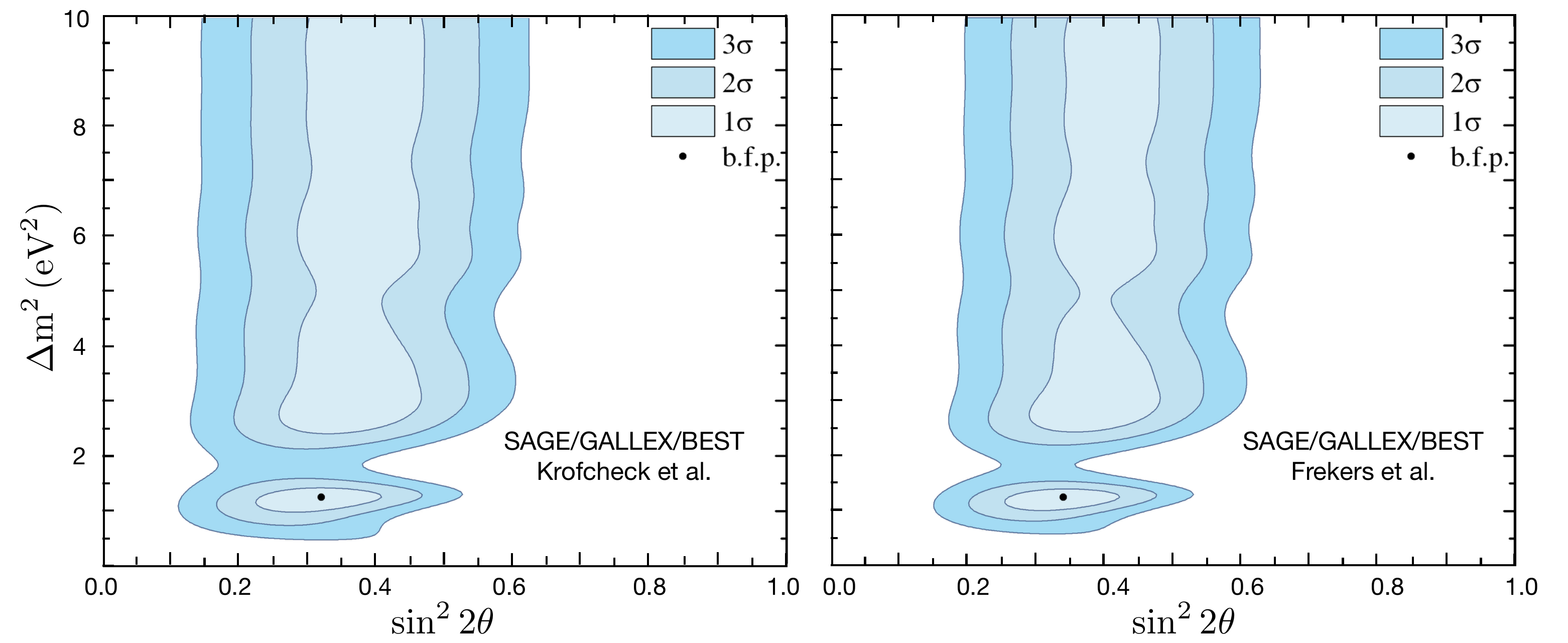}
\caption{As in Fig. 4, but combining the results of the two GALLEX and two SAGE calibrations with those of BEST.   The best-fit points
are $\mathrm{sin}^2{2 \theta} =0.32$ and $\Delta \mathrm{m}^2=1.25 $ eV$^2$ (left) and $\mathrm{sin}^2{2 \theta} =0.34$ and $\Delta \mathrm{m}^2=1.25$ eV$^2$ (right).}
\label{fig:Fig5}
\end{figure*}

Although the cross section changes found here are modest, we have updated the neutrino oscillation results of \cite{BEST1,BEST2}.
Figure \ref{fig:Fig4} gives the exclusion contours corresponding to 1$\sigma$, 2$\sigma$, and 3$\sigma$ confidence levels, using only
the BEST inner and outer results.   The cross sections used are those derived here, with the excited-state contribution extracted
from the results of \cite{Krofcheck} (left panel) and \cite{FrekersGa} (right panel).   The best-fit points correspond to
$\mathrm{sin}^2{2 \theta}  =0.41$ and $\Delta \mathrm{m}^2$=6.1 eV$^2$ and $\mathrm{sin}^2{2 \theta}  =0.45$ and $\Delta \mathrm{m}^2$=6.5 eV$^2$, respectively.
However, the chi-square space is quite shallow and flat, so solutions along a valley centered on the contours of Fig. \ref{fig:Fig4} (and Fig. \ref{fig:Fig5})
provide nearly equivalent fits.

Figure \ref{fig:Fig5} gives the exclusion contours corresponding to 1$\sigma$, 2$\sigma$, and 3$\sigma$ confidence levels, when
the BEST inner and outer results are combined with those of the two GALLEX and two SAGE calibrations.    The best-fit points correspond to
$\mathrm{sin}^2{2 \theta}  =0.32$ and $\Delta \mathrm{m}^2$=1.25 eV$^2$ and $\mathrm{sin}^2{2 \theta}  =0.34$ and $\Delta \mathrm{m}^2$=1.25 eV$^2$, respectively, for the indicated cross sections.
The shift in the best-fit results from those of Fig. \ref{fig:Fig4} reflect the shallowness of the chi-square space.

\section{Summary and Conclusions}
The published  BEST analysis employed an older $^{51}$Cr cross section from Bahcall \cite{Bahcall97}, $\sigma = \left[ 5.81^{+0.21}_{-0.16} \right] \times 10^{-45}\,\mathrm{cm}^2 \, (1 \sigma)$.  Due to the experiment's
surprising result, there is good motivation for re-examining the neutrino capture cross section on $^{71}$Ga, to determine both a more modern best value for the cross section and its uncertainty.  The latter is
particularly important in judging the significance of the BEST result.   The cross section is dominated by the transition to the ground state of $^{71}$Ge.
Since the work of \cite{Bahcall97}, changes impacting this cross section include a more accurate $Q$-value and updates in the value of $g_A$ and other weak parameters.  In addition, there are
effects such as weak magnetism and the lack of universality in radiative corrections that have not previously been evaluated quantitatively.  The first half of this paper describes these and other
corrections that typically each enter at the 0.5\% level.  We have evaluated these effects including their uncertainties, finding that they combine to yield a ground-state cross section about 2.5\% smaller than that of Bahcall \cite{Bahcall97}.\\
~\\
However, the more serious potential uncertainty is that associated with the transitions to the 175 keV $\textstyle{5 \over 2}^-$ and 500 keV $\textstyle{3 \over 2}^-$ excited states in $^{71}$Ge.  In \cite{Bahcall97} those
cross sections were taken from forward-angle (p,n) measurements by Krofcheck et al. \cite{Krofcheck}.   As was stressed in \cite{HH1,HH2} and illustrated here in Fig. \ref{fig2:BGT}, (p,n) scattering is not a reliable
probe of weak B$_\mathrm{GT}$ strengths due to the presence of a subdominant spin-tensor interaction in the effective operator for the scattering.  While normally a correction, the tensor operator can 
dominate when the competing GT
amplitude is weak. 
Haxton and Hata stressed that $^{71}$Ga is a problematic case, as the transition to the first excited state, $\textstyle{3 \over 2}^- \rightarrow \textstyle{5 \over 2}^-$ (175 keV),
is naturally associated with the $\ell$-forbidden amplitude $2p_{3/2} \rightarrow 1f_{5/2}$.  This observation was confirmed here in all three of the shell-model calculations performed.  For a pure $\ell$-forbidden 
transition, the GT amplitude is zero, while the tensor amplitude would have approximately unit strength \cite{HH1}.\\
~\\
The relationship between GT strength and (p,n) forward scattering would be restored if one could quantitatively correct for the presence of the tensor
operator.  This would be possible if one could 1) reliably determine the coefficient $\delta$ of the tensor correction (its size and uncertainty) and 2) develop some means of determining the accuracy with which the
accompanying nuclear matrix element of the tensor operator can be determined.  Despite previous work \cite{Watson,HH1,HH2} determining $\delta$ from experiment, the simple fits performed and the rather uncritical selection
of data left questions about the certainty with which $\delta$ could be established. \\
~\\
In this paper we determine $\delta$ using only measurements where experimental uncertainties have been assigned and
focusing on transitions involving relatively simple $1p$- and $2s1d$-shell nuclei where structure differences arising from the choice of shell-model effective interaction are small.  Where possible, nuclear structure differences
were quantified by exploring several effective interactions.  An attractive aspect of this approach is that an accurate evaluation of the tensor matrix element is most important in those cases where it is strong ---
and these are cases where the shell model should do well.  We see from Table \ref{tab:data} that in the four $2s1d$-shell cases where the tensor matrix element is $\approx$ 1,
the variation among the calculations is typically 10\%.  Consequently, in transitions where the GT amplitude is weak but the tensor amplitude strong, one can use theory to estimate the latter (but not the former) reliably, and thus subtract it ---
the 10\% error enters in the correction, not in end result.   With a more quantitative relationship between
(p,n) measurements and GT amplitudes thus restored, even relatively weak $M_\mathrm{GT}$ can then be extracted from the data.\\
~\\
In determining $\delta$, we found a very strong correlation between cases where $(p,n)$ and weak transition strengths disagreed,
and strong tensor matrix elements producing a ratio of $|M_\mathrm{T}/M_\mathrm{GT}|$
well above one.   Our study also underscores the fact that the weakness of excited-state contributions to the $^{71}$Ga cross section
place them in a category of transitions where important tensor corrections arise.
After folding in uncertainties from both experiment and theory, we found that $\delta = 0.074 \pm 0.008 \, (1\sigma)$.   As shown in Fig. \ref{fig2:BGT}, when the tensor 
correction is made with this value of $\delta$, excellent agreement between (p,n) cross sections and weak transition strengths is restored, even for weak GT transitions.
To the precision that $\delta$ can be determined from the available data, there is no statistical evidence for any variation with mass number:
our global fit and fits to individual transitions spanned a factor of three in mass number, from $^{13}$C to $^{39}$K. \\
~\\
As this value of $\delta$ works well for a range of $2s1d$- and $1p$-shell nuclei, the use of the same $\delta$ in the $2p1f$ shell is reasonable.  This leaves the second issue mentioned above, the need to evaluate
uncertainties associated with theory estimates of the
accompanying matrix element $M_\mathrm{T}$.  While the nuclear structure of $^{71}$Ga is more complex than that 
of the $1p$- and $2s1d$-shell nuclei used in our extraction of $\delta$, the three large-basis, full-space $2p_{3/2}1f_{5/2}2p_{1/2}1g_{9/2}$ shell-model
calculations we performed were 
reasonably consistent in their predictions of the magnitude of $M_\mathrm{T}$ and sign relative to $M_\mathrm{GT}$.  Though the spread in $M_\mathrm{T}$ is larger than found in our
$1p$- and $2s1d$-shell calculations, this spread was incorporated into a theory uncertainty that was then propagated through our analysis.  With the correction for $M_\mathrm{T}$ made, we then extracted the needed 
excited state GT strengths from the (p,n) and $(^3$He,t) results of \cite{Krofcheck} and \cite{FrekersGa}, respectively.\\
~\\
The analysis shows that $M_\mathrm{T}$ and $M_\mathrm{GT}$ interfere destructively in both excited-state transitions, which increases the $|M_\mathrm{GT}|$ extracted from experiment.  Consequently,
the excited-state contribution to the total $^{51}$Cr cross section is increased modestly, to $\approx$ 6\% and $\approx$ 9\%, depending on whether the data from \cite{Krofcheck} or \cite{FrekersGa} is used. 
The  result we obtained from the data of \cite{Krofcheck}, $\left[5.71^{+0.27}_{-0.10}\right] \times 10^{-45}$ cm$^2$ (1$\sigma$) can
be compared to the analogous result of Bahcall employed in the BEST analysis,
$\left[ 5.81^{+0.21}_{-0.16} \right] \times 10^{-45}\,\mathrm{cm}^2$ (1$\sigma$).  The results are in agreement at $1\sigma$, reflecting in part compensating changes in the present analysis due to a weaker
ground-state and stronger excited-state contributions.   

Several objections that one might have raised to the use of an older cross section --- including a more na\"ive use of the (p,n) data, absence of 
radiative and weak-magnetism corrections, and various changes in weak parameters and Coulomb corrections --- have been addressed and in combination have been found to shift the recommended
central value by only 2\%.   Most important, the analysis presented here has propagated all identified errors --- whether experimental or theoretical --- through to the end result.  Thus the error bars
reflect all known uncertainties, to the precision currently possible.   Finally, taking into account uncertainties, the extracted values of $M_\mathrm{GT}$ are in agreement with the predictions of the
shell model.  While we would certainly not advocate use of theory in estimating such weak B$_\mathrm{GT}$ strengths, nevertheless this consistency is of some comfort, as the shell model is employed
in the evaluation of the correction terms proportional to $M_\mathrm{T}$.\\
~\\
Finally, we note that a 3\% larger cross section is obtained if we base the excited-state analysis on the ($^3$He,t) data of \cite{FrekersGa}.  There would be some value in repeating both the 
(p,n) and ($^3$He,t) measurements: while the impact on the cross section is modest, the difference in the cross sections for exciting the $\textstyle{3 \over 2}^-$ state exceeds expectations, 
given the assigned error bars.\\
~\\
The lower cross section derived here from the (p,n) data of \cite{Krofcheck} slightly reduces
the size of the BEST and Ga anomalies (by $\approx$ 2\%), but certainly does not remove them.  We have demonstrated
this by repeating the sterile-neutrino oscillation analysis of \cite{BEST1,BEST2}, finding small shifts in the confidence-level contours and best-fit values for
$\mathrm{sin}^2{2 \theta}$ and $\Delta \mathrm{m}^2$.  The very well measured ground-state
transition $\sigma_{gs}$ establishes a floor on the cross section just 8\% below the value used in the BEST and earlier Ga analyses.
Even if this minimum theoretical floor were to be used --- the revised value found here is $\left[ 5.39 \pm 0.04 \right] \times 10^{-45}\,\mathrm{cm}^2$ (1$\sigma$) --- the existing
discrepancies would be reduced by about half, but not eliminated.
Furthermore, we stress that use of such an extreme minimum cross section would not be consistent with the present analysis,
as that value lies well beyond the 95\% C.L. lower bound on the total cross section derived here.\\
~\\
{\it Acknowledgments:} WCH and EJR thank Leendert Hayen and Ken McElvain for several helpful discussions.  This work was supported in part by
the US Department of Energy under grants DE-SC0004658, DE-SC0015376, and DE-AC02-05CH11231,
by the National Science Foundation under cooperative agreements 2020275 and 1630782, and by the Heising-Simons Foundation under award 00F1C7.



\noindent

\newpage


\begin{thebibliography}{60}

\bibitem{Acero2022} M. A. Acero, C. A. Arg\"uelles, N. Hostert, D. Kaira, G. Karagiorgi, et al., Snowmass white paper ``Light
Sterile Neutrino Searches and Related Phenomenology," arXiv:2203.07323 (2022)

\bibitem{Abazajian2012}  K. N. Abazajian, M. A. Acero, S. K. Agarwalla, A. A. Aguilar-Arevalo, C. H. Albright, et al.,``Light sterile neutrinos: A white
paper," arXiv:1204.5379 (2012)

\bibitem{Gariazzo2015} S. Gariazzo, C. Giunti, M. Laveder, Y. F. Li, and E. M. Zavanin, J. Phys. G: Nucl.
Part. Phys. {\bf 43}, 033001 (2015)

\bibitem{Giunti2019} C. Giunti and T. Lasserre, Annu. Rev. Nucl. Part. Sci. {\bf 69}, 163 (2019)

\bibitem{Boser2020}  S. B\"oser, C. Buck, C. Giunti, J. Lesgourgues, L. Ludhova, S. Mertens, A. Schukraft, and M. Wurm,
Prog. Part. Nucl. Phys. {\bf 111}, 103736 (2020)

\bibitem{Diaz2020}  A. Diaz, C. Arguelles, G. Collin, J. Conrad, and M. Schaevitz, Phys. Rept. {\bf 884}, 1 (2020)

\bibitem{Seo2020}  S.-H. Seo, ``Review of sterile neutrino experiments," talk presented at the 19th Lomonosov Conference, arXiv:2001.03349 (2020)

\bibitem{Dasgupta2021} B. Dasgupta and J. Kopp, Phys. Rep. {\bf 928}, 1 (2021)

\bibitem{deGouvea2022}  A. de Gouvea, G. J. Sanchez, and K. J. Kelly, arXiv:2204.09130 (2022)

\bibitem{SK} K. Abe et al. (Super-Kamiokande Collaboration), Phys. Rev. D {\bf 94}, 052010 (2016)

\bibitem{SNO} B. Aharmim et al. (SNO Collaboration), Phys. Rev. C {\bf 88}, 025501 (2013)

\bibitem{Borexino1} M. Agostini et al. (Borexino Collaboration), Ap. J. {\bf 850}, 21 (2017)

\bibitem{Borexino2} S. Appel et al., Phys. Rev. Lett. {\bf 129}, 252701 (2022)

\bibitem{review1} W. C. Haxton, R. G. H. Robertson, and A. M. Serenelli, Ann. Rev. Astron. Astrophys. {\bf 51}, 21 (2013)

\bibitem{review2} G. D. Orebi Gann, K. Zuber, D. Bemmerer, and A. Serenelli, Ann. Rev. Nucl. Part./ Phys. {\bf 71}, 491 (2021)

\bibitem{Ansel1995} GALLEX Collaboration, P. Anselmann, R. Frockenbrock, W. Hampel, G. Heusser, J. Kiko, et al., Phys. Lett. B {\bf 342}, 440 (1995)

\bibitem{Ab1999} J. N. Abdurashitov, V. N. Gavrin,  S. V. Girin, V. V. Gorbachev,  T. V. Ibragimova, A. V. Kalikhov, et al., Phys. Rev. C {\bf 59}, 2246 (1999)

\bibitem{Hampel1998} W. Hampel, G. Heusser, J. Kiko, T. Kirsten, M. Laubenstein,  et al., Phys. Lett. B {\bf 420}, 114 (1998)

\bibitem{DataSheet} B. Sigh and J. Chen, Nucl. Darta Sheets {\bf 188}, 1 (2023)

\bibitem{Ab2006} J. N. Abdurashitov,  V. N. Gavrin, S. V. Girin, V. V. Gorbachev, P. P. Gurkina, T. V. Ibragimova, et al., Phys. Rev. C {\bf 73}, 045805 (2006)

\bibitem{BEST1} V. V. Barinov,  B. T. Cleveland, S. N. Danshin, H. Ejiri, S. R. Elliott, D. Frekers, et al., Phys. Rev.Lett. {\bf 128}, 232501 (2022)

\bibitem{BEST2} V. V. Barinov, S. N. Danshin, V. N. Gavrin, V. V. Gorbachev, D. S. Gorbunov, T. V. Ibragimova, et al., Phys. Rev. C {\bf 105}, 065502 (2022)

\bibitem{HH1} N. Hata and W. C. Haxton, Phys. Lett. B {\bf 353}, 422 (1995)

\bibitem{HH2} W. C. Haxton, Phys. Lett. B {\bf 431}, 110 (1998)

\bibitem{Hampel1985} W. Hampel and L. P. Remsberg, Phys. Rev. C {\bf 31}, 666 (1985)

\bibitem{Frekers2016}  M. Allansary, D. Frekers, T. Eronen, L. Canete, J. Hakala, et al., Int. J. Mass Spectrom. {\bf 406}, 1 (2016)

\bibitem{Ab1999b} J. N. Abdurashitov, V. N. Gavrin, S. V. Girin, V. V. Gorbachev, T. V. Ibragimova, A. V. Kalikhov, et al., Phys. Rev. C {\bf 60}, 055801 (1999)

\bibitem{Benois1950} P. Benois-Gueutal, Compt. Rend. {\bf 230}, 624 (1950)

\bibitem{Odiot1956} S. Odiot and R. Daudel, J. Phys. Rad. {\bf 17}, 60 (1956)

\bibitem{Bahcall62} J. N. Bahcall,. Phys. Rev. Lett. {\bf 9}, 500 (1962)

\bibitem{Vatai70} E. Vatai, Nucl. Phys. {\bf A156}, 541 (1970)

\bibitem{Bahcall63a} J. N. Bahcall,. Phys. Rev. {\bf 131}, 1756 (1963)

\bibitem{Bahcall63b} J. N. Bahcall,. Phys. Rev. {\bf 132}, 362 (1963)

\bibitem{Bahcall65}  J. N. Bahcall, Nucl. Phys. {\bf 71}, 267 (1965)

\bibitem{Perkeo2019} B. M\"{a}rkisch, H. Mest, H. Saul, X. Wang, H. Abele, et al., Phys. Rev. Lett {\bf 122}, 242501 (2019)

\bibitem{Bahcall97} J. N. Bahcall, Phys. Rev. C {\bf 56}, 3391 (1997).

\bibitem{GCN2850} A. Gniady, E. Caurier, and F. Nowacki, unpublished

\bibitem{jj44b} B. A. Brown, unpublished; B. Cheal, E. Mane, J. Billowes, M.L. Bissell, K. Blaum, B. A. Brown, et al., Phys. Rev. Lett. {\bf 104},  252502 (2010)

\bibitem{JUN45} M. Honma, T. Otsuka, T. Mizusaki, and M. Hjorth-Jensen, Phys. Rev. C {\bf 80}, 064323 (2009)

\bibitem{JGH2012} J. G. Hirsch and P. C. Srivastava, J.Phys. G: Conference Series {\bf 387}, 012020 (2012)

\bibitem{Mane} E. Mane, B. Cheal, J Billowes, M. L. Bissell, K. Blaum, F. C. Charlwood, et al., Phys. Rev. C84, 024303 (2011)

\bibitem{Sri} P. C. Srivastava, J. Phys. G {\bf 39} 015102 (2012)

\bibitem{Klos} P. Klos, J. Menendez, D. Gazit, and A. Schwenk, Phys. Rev. D {\bf 88}, 083516 (2013)

\bibitem{Caurier} E. Caurier, J. Menendez, F. Nowacki, and A. Poves, Phys. Rev. Lett. {\bf 100}, 052503 (2008)

\bibitem{Menendez} J. Menendez, A. Poves, E. Caurier, and F. Nowacki, Phys. Rev. C {\bf 80}, 048501 (2009)

\bibitem{Johnson:2013bna}
C. W. Johnson, W. E. Ormand and P. G. Krastev, Comput. Phys. Commun. \textbf{184}, 2761 (2013)

\bibitem{Calvin} C. W. Johnson, W. E. Ormand, K. S. McElvain, and H. Shan, arXiv:1801.08432.

\bibitem{Sirlin} A. Sirlin, Rev. Mod. Phys. {\bf 50},  573 (1978)

\bibitem{Kurylov} A. Kurylov, M. J. Ramsey-Musolf, and P. Vogel, Phys. Rev. C {\bf 67}, 035502 (2003)

\bibitem{Semenov2020} S. V. Semenov, Phys. Atom. Nucl. {]bf 83}, 1549 (2020)

\bibitem{Wilkinson} D. H. Wilkinson, Nucl. Instruments Methods Phys. Res. Sec.  A  {\bf 335}, 182 (1993)

\bibitem{Hayen} L. Hayen, N. Severijns, K. Bodek, D. Rozpedzik, and X. Mougeot, Rev Mod. Phys. {\bf 90}, 015008 (2018)

\bibitem{Behrens} H.Behrens and H. J\"{a}necke, ``Numerical Tables for Beta Decay and Electron Capture,"
ed. K.-H. Hellwege (Springer-Verlag, Berlin, 1969)

\bibitem{Devries} H. De Vries, C. W. De Jaeger, and C. De Vries, At. Data Nucl. Data Tables {\bf 36}, 495 (1987)

\bibitem{Wilkinson90} D. H. Wilkinson, Nucl. Instruments Methods Phys. Res. Sec. A {\bf 290}, 509 (1990)

\bibitem{Rose} M. E. Rose, Phys. Rev. {\bf 49}, 727 (1936)

\bibitem{Czarnecki} A. Czarnecki, G. P. Lepage, and W. J. Marciano, Phys. Rev. D {\bf 61}, 073001 (2000)

\bibitem{Kuzmin1966} V. A. Kuzmin, Sov Phys. JETP {\bf 22}, 1051 (1966)

\bibitem{Bahcall1978} J. N. Bahcall, Rev. Mod. Phys. {\bf 50}, 881 (1978)

\bibitem{Baltz1984} A. J. Baltz, J. Weneser, B. A. Brown, and J. Rapaport, Phys. Rev. Lett. {\bf 53}, 2078 (1984)

\bibitem{Mathews1985} G. J. Mathews, S. D. Bloom, G. M. Fuller, and J. N. Bahcall, Phys. Rev. C {\bf 32}, 796 (1985)

\bibitem{Ikeda1964} K. Ikeda, Prog. Theor. Phys. {\bf 31}, 434 (1964); J. I. Fujita and K. Ikeda, Nucl. Phys. {\bf 67}, 143 (1965)

\bibitem{Taddeucci}  See, e.g., T.N. Taddeucci et al., Nucl. Phys. A {\bf 469}, 125 (1987)

\bibitem{Krofcheck} D. Krofcheck et al., Phys. Rev. Lett. {\bf 55}, 1051 (1985); D. Krofcheck, PhD thesis,
Ohio State University (1987) (see Table D-7)

\bibitem{FrekersGa} D. Frekers et al., Phys. Lett. B {\bf 706}, 134 (2011)

\bibitem{Watson} J. W. Watson et al., Phys. Rev. Lett. {\bf 55}, 1369 (1985)

\bibitem{Ginocchio} P. vonNeumann-Cosel and J. N. Ginocchio, Phys. Rev. C {\bf 62}, 014308 (2000)

\bibitem{Kostensalo} J. Kostensalo, J. Suhonen, C. Giunti, and P. C. Srivastava, Phys. Lett. B {\bf 795}, 542 (2019).  We believe there
is a numerical error in this work affecting the computation of $M_\mathrm{T}$.

\bibitem{Anderson83} B. D. Anderson et al., Phys. Rev. C {\bf 27}, 1387 (1983)

\bibitem{Zegers} R. G. T. Zegers, H. Akimune, S. M. Austin, D. Bazin, A. M. vandenBerg, G. P. A. Berg, et al., Phys. Rev. C {\bf 74}, 024309 (2006)

\bibitem{Grewe} E. W. Grewe, C. Baumer, A. M. van den Berg, N. Blasi, B Davids, D. De Frenne,  et al., Phys. Rev. {\bf C} 69, 064325 (2004)

\bibitem{ENSDF} See https://www.nndc.bnl.gov

\bibitem{CK} S. Cohen and D. Kurath, Nucl. Phys. {\bf 73}, 1 (1965)

\bibitem{Wildenthal} B. A. Brown and H. Wildenthal, Annu. Rev. Nucl. Part. Sci. {\bf 38}, 29 (1988)

\bibitem{USD} B. A. Brown and W. A. Richter, Phys. Rev. C {\bf 74}, 034315 (2006).

\bibitem{Frekers} D. Frekers, P. Puppe, J. H. Thies, and H. Ejiri, Nucl. Phys. A {\bf 916}, 219 (2013)

\bibitem{Barinov2018} V. Barinov, B. Cleveland, V. Gavrin, D. Gorbunov, T. Ibragimova, Phys. Rev. D {\bf 97}, 073001 (2018)

\bibitem{PPNP} S. R. Elliott, V. Gavrin, and W. C. Haxton, arXiv:2306.03299 (submitted to Prog. Part. Nucl. Phys.)

\end{thebibliography}
\end{document}